\documentclass[fleqn,usenatbib]{mnras}

\usepackage{newtxtext,newtxmath}

\usepackage[T1]{fontenc}
\usepackage{ae,aecompl}

\usepackage{graphicx}	
\usepackage{amsmath}	
\usepackage{mathrsfs}
\usepackage{subfigure}
\usepackage{placeins}
\usepackage{courier}
\usepackage{verbatim}
\usepackage{siunitx}
\usepackage{lineno}
\usepackage{lipsum}
\usepackage{paralist} 
\usepackage{tikz,xcolor,hyperref}
\usepackage{threeparttable}
\usepackage[normalem]{ulem}

\ExplSyntaxOn
  \cs_new_eq:NN \calc \fp_eval:n
\ExplSyntaxOff

\newcommand*{\formatNumber}[2][]{\num[%
  round-mode=places,
  round-precision=2,
  output-decimal-marker={.},
  #1
  ]{\calc{#2}}}

\usepackage{cleveref}
\crefname{section}{\S}{\S\S}
\Crefname{section}{\S}{\S\S}
\crefname{table}{Table}{Tables}
\Crefname{table}{Table}{Tables}
\crefname{equation}{Eqn.}{Eqns.}
\Crefname{equation}{Eqn.}{Eqns.}
\crefname{figure}{Fig.}{Figs.}
\Crefname{figure}{Fig.}{Figs.}
\crefname{paragraph}{\S}{\S\S}
\Crefname{paragraph}{\S}{\S\S}
\crefname{appendix}{Appendix}{Appendices}
\Crefname{appendix}{Appendix}{Appendices}

\definecolor{lime}{HTML}{A6CE39}
\DeclareRobustCommand{\orcidicon}{%
	\begin{tikzpicture}
	\draw[lime, fill=lime] (0,0) 
	circle [radius=0.16] 
	node[white] {{\fontfamily{qag}\selectfont \tiny ID}};
	\draw[white, fill=white] (-0.0625,0.095) 
	circle [radius=0.007];
	\end{tikzpicture}
	\hspace{-2mm}
}

\foreach \x in {A, ..., Z}{%
	\expandafter\xdef\csname orcid\x\endcsname{\noexpand\href{https://orcid.org/\csname orcidauthor\x\endcsname}{\noexpand\orcidicon}}
}

\newcommand{\Lagr}{\mathcal{L}}

\newcommand{\mul}{\mu_{l^\star}}
\newcommand{\mub}{\mu_b}
\newcommand{\mpml}{<\!\mu_{l^\star}\!>}

\newcommand{\dpml}{\sigma_{\mu_{l}^{\star}}}

\newcommand{\ks}{ {K_{s0}} }
\newcommand{\mk}{M_{K_{s0}}}
\newcommand{\mkp}{M'_{K_{s0}}}

\newcommand{\dg}{^\circ}
\newcommand{\kms   }{ \,\, \mathrm{ km  \,  s^{-1}}              }
\newcommand{\pc    }{ \,\, \mathrm{ pc  }                        }
\newcommand{\kpc   }{ \,\, \mathrm{ kpc }                        }
\newcommand{\kmskpc}{ \,\, \mathrm{ km  \,  s^{-1} \, kpc^{-1} } }

\newcommand{\masyr }{ \,\, \mathrm{ mas \, yr^{-1}}              }
\newcommand{\magn  }{ \,\, \mathrm{ mag }                        }
\newcommand{\myr   }{ \,\, \mathrm{ Myr }                        }
\newcommand{\gyr   }{ \,\, \mathrm{ Gyr }                        }

\newcommand{\degB   }{ [\mathrm{ deg }]                        }
\newcommand{\kmsB   }{ [\mathrm{ km  \,  s^{-1}}]              }

\newcommand{\kpcB   }{ [\mathrm{ kpc }]                        }
\newcommand{\kmskpcB}{ [\mathrm{ km  \,  s^{-1} \, kpc^{-1} }] }

\newcommand{\masyrB }{ [\mathrm{ mas \, yr^{-1}}]              }
\newcommand{\magnB  }{ [\mathrm{ mag }]                        }

\newcommand{\ro}{R_0}

\newcommand{\vphisun}{V_{\phi,\odot}}
\newcommand{\vpsun}{\vec{v}_{p,\odot}}

\newcommand{\Rcr}{R_\mathrm{CR}}
\newcommand{\Rolr}{R_\mathrm{OLR}}
\newcommand{\vcirc}{V_{\mathrm{circ}}}
\newcommand{\omegab}{\Omega_{\mathrm{b}}}
\newcommand{\bangle}{ \alpha_\mathrm{bar} }

\newcommand{\gravRO }{\ro = 8.2467\pm0.0093 \kpc}

\newcommand{\gravReidVphiVal}{250.63} 
\newcommand{\gravReidVphiErr}{  0.42} 
\newcommand{\GravityReidVphiNew}{\gravReidVphiVal\pm\gravReidVphiErr} 

\newcommand{\ChibaResult}{ {\dot{\Omega}}_b=-4.5\pm1.4 \kmskpc \gyr^{-1} }

\newcommand{\sgr}{ \mathrm{Sgr\,\,A^\star} }

\newcommand{\frcb}{ f_\mathrm{RC\&B} }

\newcommand{\lhood}{ \mathscr{L} }
\newcommand{\logl}{ \log_e\left( \lhood \right) } 

\newcommand{\fN}{\mathcal{N}}

\newcommand{\fU}{\mathcal{U}}

\defcitealias{portail_2017a}{P17}
\defcitealias{clarke_2019}{C19}
\defcitealias{reid_2020}{RB20}
\defcitealias{gravity_2019}{Grav2019}
\defcitealias{gravity_2020}{Grav2020}
\defcitealias{wegg_2013}{W13}
\defcitealias{wegg_2015}{W15}
\defcitealias{simion_2017}{S17}
\defcitealias{smith_2018}{S18}
\defcitealias{rattenbury_2007b}{R07}
\defcitealias{kozlowski_2006}{K06}
\defcitealias{hattori_2018b}{H18}
\defcitealias{koposov_2020}{K20}

\newcommand{\logposterior}{ \log{_e} \left[ P\left( \omegab,\, \vphisun \right) \right] }

\newcommand{\vtfiducialvalueMATH}{251.31}
\newcommand{\psfiducialvalueMATH}{33.29}
\newcommand{\vtfiducialerrorMATH}{0.20}
\newcommand{\psfiducialerrorMATH}{0.15}

\newcommand{\vtfiducial}{   \vtfiducialvalueMATH\pm\vtfiducialerrorMATH } 
\newcommand{\psfiducial}{   \psfiducialvalueMATH\pm\psfiducialerrorMATH } 

\newcommand{\vtfiducialerror}{   \vtfiducialerrorMATH } 
\newcommand{\psfiducialerror}{   \psfiducialerrorMATH } 

\newcommand{\vtWithDispMATH}{251.21}
\newcommand{\psWithDispMATH}{32.8}
\newcommand{\vtWithDispEMATH}{0.20}
\newcommand{\psWithDispEMATH}{0.13}

\newcommand{ \vtWithDisp }{   \vtWithDispMATH\pm\vtWithDispEMATH  }
\newcommand{ \psWithDisp }{   \psWithDispMATH\pm\psWithDispEMATH  }

\newcommand{\vtRCandBten}{252.17\pm0.19 } 
\newcommand{\psRCandBten}{33.97\pm0.17 } 

\newcommand{\vtRCandBtwenty}{251.7\pm0.19 } 
\newcommand{\psRCandBtwenty}{33.57\pm0.15 } 

\newcommand{\vtRCandBforty}{251.04\pm0.23 } 
\newcommand{\psRCandBforty}{33.2\pm0.18 } 

\newcommand{\vtRCandBfifty}{252.23\pm0.32 } 
\newcommand{\psRCandBfifty}{34.41\pm0.33 } 

\newcommand{\manyminCorr }{0.41 } 

\newcommand{\vtleft}{252.68\pm0.52 } 
\newcommand{\psleft}{34.39\pm0.37 } 

\newcommand{\vtright}{250.01\pm0.31 } 
\newcommand{\psright}{34.49\pm0.36 } 

\newcommand{\vtLFalphaS}{249.75\pm0.19 } 
\newcommand{\psLFalphaS}{31.84\pm0.11 } 

\newcommand{\vtLFalphaCa}{251.39\pm0.20 } 
\newcommand{\psLFalphaCa}{32.63\pm0.13 } 

\newcommand{\vtLFalphaCb}{251.91\pm0.19 } 
\newcommand{\psLFalphaCb}{32.07\pm0.13 } 

\newcommand{\vtdeltaFMATH}{ 0.39 }
\newcommand{\psdeltaFMATH}{ 0.29 }

\newcommand{\vtdeltaF}{   \vtdeltaFMATH } 
\newcommand{\psdeltaF}{   \psdeltaFMATH } 

\newcommand{\vtdeltaLFangleMATH}{1.60 }
\newcommand{\psdeltaLFangleMATH}{1.49 }

\newcommand{\vtdeltaLFangle}{   \vtdeltaLFangleMATH } 
\newcommand{\psdeltaLFangle}{   \psdeltaLFangleMATH } 

\newcommand{\vtWellipseSpiralValMATH}{251.29 } 
\newcommand{\vtWellipseSpiralErrMATH}{0.20 } 
\newcommand{\psWellipseSpiralValMATH}{33.27 } 
\newcommand{\psWellipseSpiralErrMATH}{0.15 } 

\newcommand{\vtPcontourSpiralValMATH}{252.33 }
\newcommand{\vtPcontourSpiralErrMATH}{0.32 } 
\newcommand{\psPcontourSpiralValMATH}{34.12 }
\newcommand{\psPcontourSpiralErrMATH}{0.27 }

\newcommand{\vtWellipseSpiral}{   \vtWellipseSpiralValMATH\pm\vtWellipseSpiralErrMATH }  
\newcommand{\psWellipseSpiral}{   \psWellipseSpiralValMATH\pm\psWellipseSpiralErrMATH }  
\newcommand{\vtPcontourSpiral}{   \vtPcontourSpiralValMATH\pm\vtPcontourSpiralErrMATH }  
\newcommand{\psPcontourSpiral}{   \psPcontourSpiralValMATH\pm\psPcontourSpiralErrMATH }  

\newcommand{\vtVphiPrior}{251.18\pm0.18 } 
\newcommand{\psVphiPrior}{33.25\pm0.15 } 

\newcommand{\vtVcircEilersPrior}{251.33\pm0.20 } 
\newcommand{\psVcircEilersPrior}{33.32\pm0.15 } 

\newcommand{\vtVcircReidPrior}{251.44\pm0.20 } 
\newcommand{\psVcircReidPrior}{33.55\pm0.13 } 

\newcommand{\CoRotEilers}{6.87\pm0.40}
\newcommand{\OLRaEilers}{11.28\pm0.57}
\newcommand{\OLRbEilers}{9.17\pm0.49}
\newcommand{\CoRotReid}{7.11\pm0.38}
\newcommand{\OLRaReid}{11.88\pm0.53}
\newcommand{\OLRbReid}{9.37\pm0.57}

\newcommand{\vtdeltaDispMATH}{ abs(\vtWithDispMATH-\vtfiducialvalueMATH) }  
\newcommand{\psdeltaDispMATH}{ abs(\psWithDispMATH-\psfiducialvalueMATH) } 

\newcommand{\vtdeltaDisp}{   \formatNumber{ \vtdeltaDispMATH } }  
\newcommand{\psdeltaDisp}{   \formatNumber{ \psdeltaDispMATH } }

\newcommand{\vtdeltaSpiralMATH}{ abs(\vtPcontourSpiralValMATH-\vtfiducialvalueMATH) } 
\newcommand{\psdeltaSpiralMATH}{ abs(\psPcontourSpiralValMATH-\psfiducialvalueMATH) } 

\newcommand{\vtdeltaSpiral}{   \formatNumber{ \vtdeltaSpiralMATH }  } 
\newcommand{\psdeltaSpiral}{   \formatNumber{ \psdeltaSpiralMATH }  } 

\newcommand{\vtfinalerrorMATH}{ \formatNumber{ sqrt( \vtfiducialerrorMATH*\vtfiducialerrorMATH +\vtdeltaDispMATH*\vtdeltaDispMATH + \vtdeltaFMATH*\vtdeltaFMATH + \vtdeltaLFangleMATH*\vtdeltaLFangleMATH + \vtdeltaSpiralMATH*\vtdeltaSpiralMATH ) }}

\newcommand{\psfinalerrorMATH}{ \formatNumber{ sqrt(\psfiducialerrorMATH*\psfiducialerrorMATH+\vtdeltaDispMATH*\vtdeltaDispMATH+ \psdeltaDispMATH*\psdeltaDispMATH + \psdeltaFMATH*\psdeltaFMATH + \psdeltaLFangleMATH*\psdeltaLFangleMATH + \psdeltaSpiralMATH*\psdeltaSpiralMATH) } }

\newcommand{  \psfinal  }{  \psfiducialvalueMATH\pm\psfinalerrorMATH} 
\newcommand{  \vtfinal  }{  \vtfiducialvalueMATH\pm\vtfinalerrorMATH} 

\title[The Milky Way Bar Pattern Speed]{The Pattern Speed of the Milky Way Bar/Bulge from VIRAC \& \textit{Gaia}}

\author[Clarke \textit{\&} Gerhard]{
Jonathan P. Clarke$^{1}$\thanks{E-mail: jclarke@mpe.mpg.de}\orcidA{} and
Ortwin Gerhard$^{1}$\thanks{E-mail: gerhard@mpe.mpg.de}\orcidB{}
\\
$^{1}$ Max-Planck-Institut f{\"u}r Extraterrestrische Physik, Gie{\ss}enbachstra{\ss}e, D-85748 Garching, Germany 
}

\date{Accepted XXX. Received YYY; in original form ZZZ}

\pubyear{2022}

\newif\ifTHESIS
\THESISfalse

\begin{document}
\label{firstpage}
\pagerange{\pageref{firstpage}--\pageref{lastpage}}
\maketitle

\begin{abstract}

We compare distance resolved, absolute proper motions in the Milky Way bar/bulge region to a grid of made-to-measure dynamical models with well defined pattern speeds. The data are obtained by combining the relative VVV Infrared Astrometric Catalog v1 proper motions with the \textit{Gaia} DR2 absolute reference frame.
We undertake a comprehensive analysis of the various errors in our comparison, from both the data and the models, and allow for additional, unknown, contributions by using an outlier-tolerant likelihood function to evaluate the best fitting model.
We quantify systematic effects such as the region of data included in the comparison, the possible overlap from spiral arms, and the choice of synthetic luminosity function and bar angle used to predict the data from the models. Resulting variations in the best-fit parameters are included in their final errors.
We thus measure the bar pattern speed to be $\omegab=\psfinal\kmskpc$ and the azimuthal solar velocity to be $\vphisun=\vtfinal\kms$. These values, when combined with recent measurements of the Galactic rotation curve, yield the distance of corotation, $6.5 < R_\mathrm{CR}\,\kpcB < 7.5$, the outer Lindblad resonance (OLR), $10.7 < R_\mathrm{OLR}\,\kpcB < 12.4$, and the higher order, $m=4$, OLR, $8.7 < R_\mathrm{OLR_4}\,\kpcB < 10.0$.
The measured pattern speed provides strong evidence for the "long-slow" bar scenario.
\end{abstract}




\section{Introduction}\label{c22a:sec:intro_c22a}

\subsection{The Pattern Speed, \texorpdfstring{$\omegab$}{OmegaB}, of the Milky Way Bar}

The Milky Way (MW) bulge is dominated by a triaxial bar structure \citep{lopez_corredoira_2005,rattenbury_2007a,saito_2011,wegg_2013}.
Understanding the structure and dynamics of the Galactic bar and bulge is essential for interpreting a wide variety of MW bar/bulge observations including:
\begin{inparaenum}
\item the X-shape \citep{nataf_2010,mcwilliam_2010} and its kinematics \citep{gardner_2014,williams_2016} in the boxy/peanut (b/p) bulge \citep{wegg_2013,li_zhaoyu_2015};
\item the high line-of-sight (LOS) velocity peaks observed in the bulge \citep{nidever_2012,molloy_2015,zhou_2021};
\item the quadrupole patterns seen in VIRAC/\textit{Gaia} proper motion correlations \citep{clarke_2019};
\item the vertex deviation in the bulge \citep{babusiaux_2010,sanders_2019a,simion_2021}; and
\item the kinematics of the stellar populations in the long bar \citep{bovy_2019,wegg_2019b,wylie_2022a}.
\end{inparaenum}

An essential parameter for characterising the bar is the pattern speed, $\omegab$, which directly influences the bar length \citep[e.g.][$\approx 4.6\pm0.3\kpc$ for their "thin" long bar]{wegg_2015}, as bar supporting orbits cannot exist far beyond corotation \citep{contopoulos_1980,aguerri_1998}.
Using bulge stellar kinematics \citet[hereafter P17]{portail_2017a} estimated $\omegab=39\pm3.5\kmskpc$ by modelling several MW bulge surveys. This result was confirmed through application of the \citet{tremaine_1984a} method to VVV/VIRAC data \citep{sanders_2019b}, and by applying the continuity equation to APOGEE data \citep{bovy_2019}.

The bar drives the dynamics of gas in the inner Galaxy, generating strong non-circular motions \citep[e.g.,][]{binney_1991}.
There have been many attempts using hydrodynamical models to match the observed gas kinematics in the MW \citep[][]{ englmaier_1999,fux_1999, baba_2010,sormani_2015a,pettitt_2020} using various potentials and spiral/bar components.
$\omegab$ sets the resonant radii at which the gas flow transitions between orbit families meaning that a realistic model of the gas can place strong constraints on this parameter.
While some older studies have reported rather high values, $50\!< \!\omegab \, \kmskpcB \!<\!60$, \citep[fast-short bar, e.g.,][]{fux_1999,debattista_2002,bissantz_2003} more recent works have determined lower values, $33 \!<\! \omegab \, \kmskpcB \!<\! 40$ \citep[long-slow bar,][]{sormani_2015b,li_zhi_2016,li_zhi_2022}.

The bar also shapes the disk kinematics through resonances. A classic example is the \textit{Hercules} stream, modelled originally as the Outer Lindblad resonance (OLR) of a $50<\omegab \,\,\kmskpcB <60$ bar \citep[e.g.][]{dehnen_2000,minchev_2010,antoja_2014} but more recently, as the corotation resonance (CR) \citep{perez_villegas_2017,monari_2019b,chiba_2021b} or 4:1/5:1 OLR of a long-slow bar \citep{hunt_2018b,asano_2020}.
Bar resonances and/or spiral arms are also likely to explain the multiple structures seen by \textit{Gaia} in the extended solar neighbourhood (SNd) \citep[][Fig. 22]{katz_2018_gaia}.
While some analyses favour short-fast bar models \citep[e.g.,][]{fragkoudi_2019} or steady spiral patterns \citep[e.g.,][]{barros_2020}, most favour transient spiral arms \citep{hunt_2018c,sellwood_2019} and a long-slow bar \citep{monari_2019a, khoperskov_2020,binney_2020, kawata_2021,trick_2022}.
The effects of the bar and spiral arms are difficult to disentangle \citep{hunt_2019} emphasising the need for accurate, independent measurements of $\omegab$.

The bar's influence even stretches beyond the bulge and disk and into the stellar halo. One example being the truncation of Palomar 5 due to the different torques exerted on stars as the bar sweeps past \citep{pearson_2017,banik_2019,bonaca_2020}.

Bars can slow down over time as angular momentum is transferred to the dark matter halo \citep{debattista_2000,valenzuela_2003,martinez_valpuesta_2006}.
Conversely, they can also gain angular momentum as they channel gas towards the GC \citep{van_albada_1982,regan_2004}.
However only recently \citet{chiba_2021a} considered the effect of a \textit{decelerating} bar on local stellar kinematics. Their model reproduced \textit{Hercules} with its CR resonance and dragging by the slowing bar generated multiple resonant ridges found in action coordinates. Perhaps most importantly they thereby showed that models using a constant $\omegab$ can lead to incorrect conclusions.
The dynamical effects of the bar are further complicated as $\omegab$ might vary by as much as $20\%$ on a timescale of $60$ - $200$ Myr due to interactions between spiral structure and the bar \citep{hilmi_2020} although these values may be model dependent.

The first step to tackling these more complex effects is to better understand the current $\omegab$ value.
In this work we provide a robust measurement, from the inner bar/bulge, in excellent agreement with recent studies that used data from the SNd \citep{binney_2020,chiba_2021b}.

\subsection{The Tangential Solar Velocity, \texorpdfstring{$\vphisun$}{VphiSun}}

To move past a heliocentric view of the MW requires precise knowledge of the sun's motion within the MW. 
A recent, high precision measurement of $\vphisun$ combined the \citet[][hereafter Grav2020]{gravity_2020} measurement of $\gravRO$ with the proper motion of $\sgr$ from Very Long Baseline Array radio observations \citep[hereafter RB2020]{reid_2020}. Assuming $\sgr$ is at rest with respect to the centre of the bulge and disk, the longitudinal (latitudinal) proper motion can be converted to the azimuthal (vertical) solar velocity with $v_{i,\odot}\left( \!\!\kms \right)=-4.74 \cdot \mu_{i,A*}\left( \!\!\masyr \right) \cdot \ro\left( \!\!\kpc \right)$, resulting in a total solar tangential velocity $\vphisun=\GravityReidVphiNew\kms$.
Consistent measurements were made using a newly discovered hypervelocity star \citep{koposov_2020} and using the solar system's acceleration from \textit{Gaia} EDR3 data \citep{bovy_2020}.

Here we use the kpc-scale bulge rather than $\sgr$ to obtain a precise measurement of $\vphisun$. Whether these two approaches give consistent answers provides information on whether both components are at rest relative to each other.

In an axisymmetric galaxy the local standard of rest (LSR) is defined as a circular orbit through the solar position, with velocity $\vec{v}_\mathrm{LSR} = (0,\, \vcirc(\ro),\,0)$ \citep{galactic_dynamics}. The solar peculiar motion, or its negative, the velocity of the LSR relative to the sun, is found by considering the streaming velocities of samples of nearby stars with different velocity dispersions and extrapolating to small dispersion. In this case, $\vphisun$ is  the combination of the circular velocity, $\vcirc(\ro),$ and the tangential component, $V_\odot$, of the solar peculiar velocity.
\footnote{The solar peculiar velocity vector, relative to the LSR, is here defined as $\vpsun=(U,V,W)_\odot$ where $U_\odot$ is radially inwards, $V_{\odot}$ is tangential in the direction of Galactic rotation, and $W_\odot$ is vertically upwards.}

However, in the MW's bar+spiral gravitational potential, where stars near the sun are no longer on families of perturbed circular orbits, the definition of the LSR is more complicated. It is still useful to define an average circular velocity, $\vcirc(\ro)$ at $\ro$, as the angular velocity of a fictitious circular orbit in the azimuthally averaged potential \citep[sometimes called the rotational standard of rest, RSR, see] []{shuter_1982,bovy_2012f}. However due to the non-axisymmetric perturbations we now expect systematic streaming velocities relative to this RSR
\footnote{As a simple} example, the zero-dispersion LSR for stars on dynamically cold, non-resonant orbits in a weakly barred potential between corotation and the OLR would be a near-elliptical closed orbit with faster (slower) tangential velocity than $\vcirc(\ro)$ on the bar's major (minor) axis.. The velocity maps presented by \citet[][e.g. their Fig. 10]{katz_2018_gaia} show a complicated streaming velocity field in the nearby disk. In such cases, the LSR as determined from local star kinematics will not, in general, coincide with the globally averaged RSR circular velocity \citep{Drimmel_2018}, i.e., $V_\odot$ is measured relative to an LSR that will itself have a non-circular velocity with respect to the RSR, $\vec{v}_{\mathrm{LSR}}=\left(U,\,V,\,W\right)_\mathrm{LSR}$, such that the total azimuthal LSR velocity $V_{\phi,\mathrm{LSR}}= \vcirc(\ro)+ V_\mathrm{LSR}$ and,
\begin{linenomath}\begin{equation}
  \vphisun = \vcirc(\ro)+ V_\mathrm{LSR}+V_\odot.
\end{equation}\end{linenomath}
Multiple studies have constrained individual or combined velocity components in eq.~1 (see \citet{bland_hawthorn_2016} for an overview):
$\vcirc(\ro)$ has been determined using stellar streams \citep{koposov_2010,kupper_2015,malhan_2020}, LOS velocities from APOGEE \citep{bovy_2012f}, MW mass modelling \citep{mcmillan_2017}, cepheids in \textit{Gaia} DR2 \citep{kawata_2019}, red giants stars with precise parallax \citep{eilers_2019}, and parallaxes and proper motions of masers \citep{reid_2019}.
Standard values for $\vpsun$ were published by \citet{schoenrich_2010} although it has been measured many times \citep[e.g.][]{delhaye_1965,dehnen_1998,binney_2010}.
Not accounting for the additional $V_\mathrm{LSR}$ term can lead to apparently contradictory measurements and care should be taken when combining measurements from different sources. Accurate measurements of $\vphisun$, $\vcirc(\ro)$, and $V_\odot$ potentially constrain $V_\mathrm{LSR}$.

\subsection{Our Approach}

The VVV InfraRed Astrometric Catalogue (VIRAC) \citep{smith_2018} contains $\approx 1.75\times10^8$ proper motions across the Galactic bulge region, roughly ($-10 < l \,\,\degB < 10$, $-10 < b \,\,\degB < 5$). When combined with \textit{Gaia} data \citep{brown_2018_gaia} to provide the absolute reference frame these data provide an extraordinary opportunity to study the kinematics through the bulge region.
Using various radial velocity and stellar density information in the bulge \citetalias{portail_2017a} constructed a grid of dynamical models, with well defined $\omegab$ values, using the made-to-measure (M2M) method.
These models are a powerful resource because, unlike many other dynamical models, they have been iteratively adapted to fit observed star counts and kinematics, providing superior parity between model and observations.
Kinematic maps of the VIRAC/\textit{Gaia} (gVIRAC, see \cref{c22a:subsec:gVIRAC}) data and the $\omegab=37.5 \kmskpc$ M2M dynamical model \citepalias{portail_2017a} were \textit{qualitatively} compared in \citet[][hereafter C19]{clarke_2019} finding excellent agreement despite the models not having been fit to the data.

The purpose of this paper is to provide accurate measurements of $\omegab$ and $\vphisun$.
We shall utilise the \citetalias{portail_2017a} M2M models for a systematic, quantitative comparison to the gVIRAC data.
We further derive CR and OLR distances from the GC assuming recently determined Galactic rotation curves \citep{eilers_2019,reid_2019}.
The structure of the paper is as follows. 
In \cref{c22a:sec:models_data} we present the data and models we are comparing. 
\cref{c22a:sec:errors} describes the analysis of the sources of error in our comparison and \cref{c22a:sec:statistics} outlines our adopted approach for measuring $\omegab$ robustly.
In \cref{c22a:sec:testing} we present tests carried out to ensure the results are also robust against known systematics (choice of luminosity function and bar angle, effect of spiral arms, and region in the inner bar/bulge where we make the measurement).
\cref{c22a:sec:corrotation} describes the inferred resonant radii in the disk. Finally, we discuss the results in a wider context in \cref{c22a:sec:discussion} and summarise and conclude in \cref{c22a:sec:conclusion}.


\section{Models \& Data}\label{c22a:sec:models_data}
In this section we will describe the data we are using, a combination of VIRAC and \textit{Gaia}, the M2M models constructed in \citetalias{portail_2017a}, and the methods used to predict the VIRAC/\textit{Gaia} data from the models. The section ends with a description of the simple masking approach we take to exclude less robust kinematic data from the comparison.

\subsection{VIRAC + Gaia: gVIRAC}\label{c22a:subsec:gVIRAC}

VIRACv1 \citep{smith_2018}; a catalogue of 312 587 642 unique, albeit \textit{relative}, proper motion measurements covering 560 $\mathrm{deg}^2$ of the MW southern disc and bulge derived from the VVV survey \citep{minniti_2010}. 
The bulge observations consist of a total of 196 separate tiles spanning $-10<l\,[\mathrm{deg}]\,<10$ and $-10<b\,[\mathrm{deg}]\,<5$. Each tile has a coverage of $\approx 1.4\dg$ in $l$ and $\approx 1.1\dg$ in $b$ and is observed for 50 to 80 epochs from 2010 to 2015. Typical errors are $\approx 0.7\masyr$ for brighter stars away from the Galactic plane but can be as large as $>1.2\masyr$ for fainter, more in-plane stars.

The following summarises the extraction of a red giant branch (RGB) star sample in the MW bulge/bar region (see \citetalias{clarke_2019} for a detailed discussion).
\begin{inparaenum}
\item VIRAC provides relative proper motions. Absolute proper motions were obtained by cross-matching to \textit{Gaia's} DR2 absolute reference frame \citep{lindegren_2018_gaia}. gVIRAC is used here to refer to this combination of VIRAC and \textit{Gaia} data.\footnote{The upcoming release of VIRACv2 (Smith et al. in preparation) will contain the proper motions determined from improved photometry and will be calibrated to \textit{Gaia} EDR3 \citep{brown_2021_gaia}.}
\item RGB stars in the bulge were distinguished from foreground main sequence stars according to a Gaussian mixture model of the $(H-K_{s})$ vs $(J-K_{s})$ distribution.
\item Magnitudes were extinction corrected using the extinction map of \citet{gonzalez_2012} and the \citet{nishiyama_2009} coefficients.
\end{inparaenum}

At this point the RGB stars have been separated from foreground main sequence stars however, due to the large range in RGB absolute magnitudes, each apparent magnitude interval is composed of stars spanning a large physical distance range.
The red clump (RC) can be used as a standard candle due to the narrowness of its intrinsic luminosity function \citep{stanek_1994}. 
The RC is not easy to extract cleanly; there are no definitive photometric measures by which to separate it from other RGB stars. Therefore the red clump \& bumps (RC\&B) population, the combination of the RC, the red giant branch bump (RGBB), and the asymptotic giant branch bump (AGBB), is used which is much easier to isolate (see \cref{c22a:tab:acronyms_c22a} for a summary of stellar type acronyms used in this paper). 
The RGBB + AGBB contamination fraction was measured by \citet{nataf_2011} to be $24\%$.
The RC\&B population sits on top of the smooth exponential continuum \citep{nataf_2010} which we refer to as the red giant branch continuum (RGBC).

The RGBC velocity distribution was measured, independent of the RC\&B, at $14.1 \leq \ks\,\, \magnB \leq14.3$, where there is little to no contamination by the brighter RC\&B.
Subtracting the RGBC velocity distribution at brighter magnitude intervals, suitably scaled according to the observed RGBC luminosity function, allows the kinematics of just the RC\&B, for which magnitude is a proxy for distance, to be measured.
The individual magnitude intervals used here have width $\Delta K_{s0}=0.1$ mag, and for brevity we shall refer to them, across all VIRAC tiles, as \textit{voxels}, $(l_i, b_j, K_{s0,k})$.

For our later analysis we remove the two most in-plane rows of tiles from the analysis as they are affected by extinction and crowding rendering the RC\&B kinematic measurements untrustworthy. Additionally we only consider longitudinal proper motions which are far more sensitive to $\omegab$ and $\vphisun$ for our quantitative comparison.

\begin{table}
	\centering
	\caption[List of stellar type acronyms.]{
	List of stellar type acronyms for reference. The bottom two rows represent composite groups of the initial four.
	}
	\label{c22a:tab:acronyms_c22a}
	\begin{tabular}{ll}
		\hline\\[-5pt]
		Acronym & Definition \\[2pt]
		\hline\\[-5pt]
		RC    & Red clump \\[2pt]
		RGBB  & Red giant branch bump \\[2pt]
		AGBB  & Asymptotic giant branch bump \\[2pt]
		RGBC  & Red giant branch continuum \\[6pt]
		RGB   & Red giant branch \\[2pt]
		RC\&B & Red clump and bumps\\[2pt]
		\hline
	\end{tabular}
\end{table}

\begin{table}
	\centering
	\caption[Parameters of the \citet{wegg_2013} luminosity function.]{
	Parameters of the \citetalias{wegg_2013} luminosity function. The mean and dispersion of the RC and RGBB gaussians are set to the values quoted in \citetalias{wegg_2013}.
	}
	\label{c22a:tab_w13_lf}
	\begin{tabular}{l S[table-format=2.4] @{\hspace{1.5cm}} l S[table-format=2.4]}
		\hline\\[-5pt]
		{Parameter} & {Value} & {Parameter} & {Value} \\[2pt]
		\hline\\[-5pt]
		$A_{\mathrm{RGBC}}$      &  0.1577 & $A_{\mathrm{RGBB}}$      &  0.0362 \\[2pt]
		$\alpha$                 &  0.7302 & $\mu_{\mathrm{RGBB}}$    & -0.91   \\[2pt]
		$\beta$                  &  0.0305 & $\sigma_{\mathrm{RGBB}}$ &  0.19   \\[6pt]
		$A_{\mathrm{RC}}$        &  0.1456 & $A_{\mathrm{AGBB}}$      &  0.0122 \\[2pt]
		$\mu_{\mathrm{RC}}$      & -1.72   & $\mu_{\mathrm{AGBB}}$    & -3.2126 \\[2pt]
		$\sigma_{\mathrm{RC}}$   &  0.18   & $\sigma_{\mathrm{AGBB}}$ &  0.3488 \\[2pt]
		\hline
	\end{tabular}
\end{table}

\subsection{M2M Dynamical Models}

We will compare the VIRAC data with the predictions of the M2M barred dynamical models of the Galactic bulge obtained by \citetalias{portail_2017a}. 
The M2M models were constructed by gradually adapting dynamical N-body models to fit the following constraints:
\begin{inparaenum}
\item the density of RC stars in the bulge region computed by deconvolution of VVV RC + RGBB luminosity functions \citep[][hereafter WG13]{wegg_2013};
\item the magnitude distribution in the long bar determined by \citet[][hereafter W15]{wegg_2015} from UKIDSS \citep{lucas_2008} and 2MASS \citep{skrutskie_2006}; and
\item stellar radial velocity measurements from the BRAVA \citep{howard_2008,kunder_2012} and ARGOS \citep{freeman_2013,ness_2013} surveys.
\end{inparaenum}
We note that these models assume a single disk beyond the bulge region and do not include a separate thick disk component.

We consider a sequence of models from P17 with well determined $\omegab$ in the range $30.0$ to $45.0 \kmskpc$.
For each $\omegab$ we select their model with the overall best mass-to-clump ratio, $M/n_\mathrm{RC}=1000$. The extra central mass, $M_c$, that \citetalias{portail_2017a} required in addition to the stellar bar/bulge is chosen for each $\omegab$ to minimise the $\chi^2$ of the stellar density and total rotation curve obtained by \citetalias{portail_2017a}. We omit the kinematic constraints used by \citetalias{portail_2017a} in this evaluation because the gVIRAC data to which we compare the models result in much stronger constraints on the bulge kinematics. We include the density so that the models, when re-convolved, are best able to reproduce the gVIRAC data, and the rotation curve constraint to optimise the dark matter halo. We thus find that, for all $\omegab$, the model with $M_c=10^9 M_\odot$ is preferred. This is in good agreement with the Nuclear Stellar Disk mass determined recently by \citet{sormani_2022}. We have also verified that the corresponding models match the gVIRAC velocity dispersion maps better than models with larger $M_c$.
%

\subsection{Predicting the gVIRAC Kinematics}\label{c22a:subsec:predicting_gVIRAC}

\citetalias{wegg_2013} used the \texttt{BASTI} isochrones to construct a synthetic luminosity function (synth-LF) for the bulge RGB stars of a 10 $\gyr$ old stellar population. 
This synth-LF was used to deconvolve line-of-sight (LOS) observed luminosity functions (obs-LF) from VVV to produce 3D RC density maps.\footnote{
We make the distinction between synth-LF, true-LF, and obs-LF as they are three distinct concepts that are all commonly called `LFs'. A synth-LF is generated for simulations, using isochrones, an initial mass function, and a metallicity distribution, and is an approximation to the true absolute magnitude distribution of a given stellar population; the true-LF. An obs-LF is a function of apparent magnitude and represents the convolution of a synth-LF with a LOS density distribution.}

The \citetalias{wegg_2013} synth-LF has 4 components corresponding to different stages of stellar evolution.
There is a near-exponential background for the RGBC given by,
\begin{linenomath}\begin{equation}\label{c22a:eqn_lf_rgbc}
    \Lagr_{\mathrm{RGBC}}\left( \mkp \right) = A_{\mathrm{RGBC}} \exp{ \left( \alpha \mkp + \beta {\mkp}^2 \right)},
\end{equation}\end{linenomath}
and separate gaussian components for each of the RC, RGBB, and AGBB,
\begin{linenomath}\begin{equation}\label{c22a:eqn_lf_gauss}
    \Lagr_{i}\left( \mkp \right) = \frac{A_i}{\sqrt{2 \pi \sigma_i^2}} \exp{ \left( -\frac{1}{2} \left(\frac{\mkp - \mu_i}{\sigma_i}\right)^2  \right) },
\end{equation}\end{linenomath}
where $i$ denotes the stellar population component. 
Parameter values are given in \cref{c22a:tab_w13_lf}.
The RC density measurements of \citetalias{wegg_2013} were computed assuming $\ro=8.3\kpc$. We shift the synth-LF taking $\mkp=\mk-0.026$ to account for the more recent $\ro=8.2\kpc$ GC distance \citep{bland_hawthorn_2016,gravity_2019}.
We also allow for a shift in RC absolute magnitude, due to the vertical metallicity gradient in the bulge, by adding a further, $z$-dependent shift to the synth-LF magnitudes, see \cref{c22a:appendix:metallicity}.
The deconvolution process produces a LOS density profile with a systematic error introduced by any differences between the synth-LF and the true-LF. 
For a given apparent magnitude distribution using a broader-than-reality LF will result in a narrower-than-reality density distribution and vice versa.
\citetalias{portail_2017a} fitted the grid of M2M models to these 3D density maps.
Reconvolving the model density distribution with a different synth-LF will introduce further systematic errors compounding the effect.

Therefore, when predicting the gVIRAC kinematics, we take the \citetalias{wegg_2013} synth-LF as our fiducial assumption but will estimate the systematic effects of varying the synth-LF in \cref{c22a:subsec:test_LF_alpha}.
Each model particle is treated as a stellar population according to the synth-LF. For a given apparent magnitude interval, the particle's contribution is obtained by shifting the synth-LF according to the particle's distance modulus and integrating over the bin width (see \citetalias{clarke_2019} for a detailed description).
When computing proper motion dispersions for the particles in a given apparent mag interval we allow for the broadening effect of proper motion measurement errors on the dispersion measurements by adding an appropriate Gaussian random deviation to each individual model proper motion (see \cref{c22a:subsubsec:gVIRAC_broad}).


\begin{figure*}
    \ifTHESIS
        \includegraphics[width=\textwidth]{Figures_c22a/RCB2ALL_ratio_Regions_ForThesis.pdf}
    \else
    	\includegraphics[width=\textwidth]{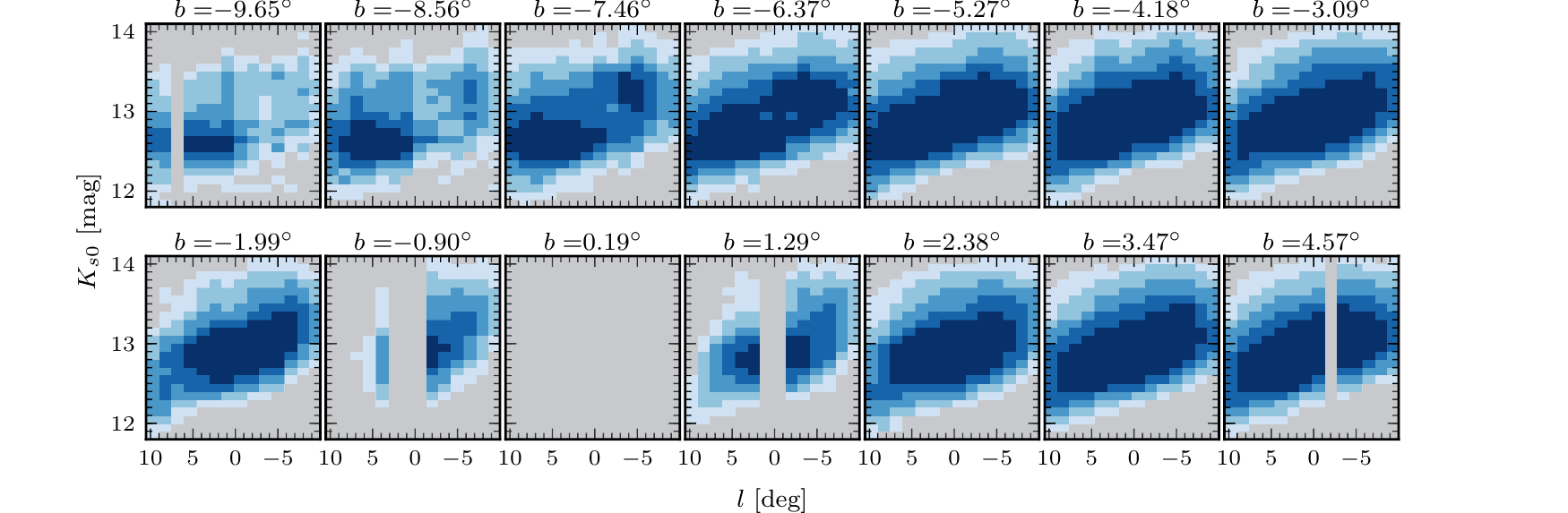}
    \fi
    \caption[Map of the RC\&B star fraction throughout the Galactic bulge volume.]{
    Map showing the RC\&B fraction of all stars present as a function of magnitude and according to the VVV tiling pattern. The white region outlines all fields in which the RC\&B comprise at least $10\%$ of all stars in the magnitude interval. The darkest blue shows where the RC\&Bs account for at least $50\%$ and the intermediate colours represent $20\%$, $30\%$, and $40\%$.
    We see the split RC effect first shown by \citet{nataf_2010,mcwilliam_2010} in the $b=-6.37\dg$ panel where there are two peaks along the $l \approx 0\dg$ lines of sight. Furthermore we see the orientation of the bar with the near end at positive longitude from the tilt of the outlined regions.
    The vertical grey stripes in the extreme $b$ panels are due to a lack of colour information preventing the extraction of the RGB stars. The vertical stripes near the Galactic plane are where completeness prevents us from fitting the RGBC (necessary for extracting the RC\&B). 
    }
    
    \label{c22a:fig:RCBs_fractions}
\end{figure*}

\subsection{Importance of the Red Clump Fraction in the Bulge}\label{c22a:subsec:RCBfrac}

RC stars in the barred bulge cause a peak in the observed magnitude distribution at $\ks \approx 12.8 \magn$ although this varies with longitude due to the bar orientation.
The peak is relatively narrow, $\Delta\ks \approx 1 \magn$, due to the localised high density of the bulge and the intrinsically narrow RC true-LF. 
In contrast, the RGBC at a given distance is far more broadly distributed in magnitude, hence its removal as discussed in \cref{c22a:subsec:gVIRAC}.
\cref{c22a:fig:RCBs_fractions} shows the RC\&B fraction, $\frcb$, as a function of magnitude in horizontal slices through the bulge.
White areas indicate regions where the RC\&B contributes $>10\%$ of the stars in the magnitude interval.
The darkest blue shows where the RC\&B comprises $>50\%$ and intermediate colours represent $>20\%$, $>30\%$, and $>40\%$ fractions.
We see the split RC \citep{nataf_2010,mcwilliam_2010} prominently in the $b=-6.37\dg$ panel; the magnitude distribution peaks twice along the $l\approx 0\dg$ LOS.
The orientation of the bar, with the near end at positive longitude, is also obvious.

The majority of the data to which the \citetalias{portail_2017a} models have been fit is distance resolved RC data.
Thus, the regions in magnitude space that have a larger contribution from RC\&B stars are better constrained than regions with smaller contributions. 
Thus there is a question as to exactly which data we consider in our analysis.
Using too strong a $\frcb$ criteria will remove a large amount of usable data while we find the $>10\%$ case includes a disproportionate number of voxels with larger systematic errors compared to stricter selections (see \cref{c22a:sec:errors}).
We therefore take the $\frcb>30\%$ criteria as our fiducial assumption and we test the effect of this choice in \cref{c22a:subsec:test_frcb}.
At high latitude, $|b|>7\dg$, the $\frcb$ map becomes noisy; this is a direct result of noise in the VVV obs-LFs which, when compared to the  RGBC exponential fit, shifts the inferred $\frcb$ above and below the thresholds.

\section{Error Analysis}\label{c22a:sec:errors}

An essential part of a quantitative model-to-data comparison is a thorough analysis of the possible sources of error in both models and data. 
In this section we discuss the statistical and systematic uncertainties we consider and describe the methods used for estimating these errors.  
Readers who are primarily interested in the results can go directly to \cref{c22a:fig:vvv_err_dists,c22a:fig:mod_err_dists}, which show the various error distributions for the gVIRAC data, and the M2M models, respectively.

\subsection{Sources of Uncertainty in gVIRAC}

\subsubsection{VIRAC Broadening: Proper Motion Errors}\label{c22a:subsubsec:gVIRAC_broad}

There is an uncertainty in the observed dispersions intrinsic to the VIRAC data itself.
Each gVIRAC proper motion measurement has a corresponding Gaussian-distributed uncertainty. 
These individual proper motion uncertainties are not equal within a given ($l$, $b$, $\ks$) voxel but have a peak and then a long tail towards larger errors.
The peak error varies from $\approx 0.7 \masyr$ at high latitude and bright magnitudes but can become as large as 1.2 to 1.4 $\masyr$ at lower latitudes and fainter magnitudes.

These errors broaden the true proper motion distribution, $N_\mathrm{true}\left(\mul\right)$, such that the observed dispersion $\sigma_\mathrm{obs}$ in a $(l,\,b,\,\ks)$ voxel becomes larger than the true dispersion, $\sigma_\mathrm{true}$. 
To take this into account, we use the following simplified approach. First we approximate the error distribution in the $i^\mathrm{th}$ voxel by a single value, the median proper motion error, $\epsilon_{i}$, and broaden the model dispersion by adding a Gaussian random deviation to each particle's proper motion, $\mu_i \xrightarrow[]{} \mu_i + \fN\left( 0, \, \epsilon_i \right)$. This correctly convolves the non-Gaussian proper motion distribution with the median error however we include an additional error on the observed $\dpml$ defined by 
\begin{linenomath}\begin{equation}
    \delta_\sigma 
    \triangleq 
    \sigma_\mathrm{obs} - \sigma_\mathrm{true} 
    =
    \sigma_\mathrm{obs} - \sqrt{ \sigma_\mathrm{obs}^2 - \epsilon^2 },
\end{equation}\end{linenomath}
to accommodate the uncertainty in approximating the error distribution by the median value.
The mean $\mpml$ are unaffected by this convolution.

\subsubsection{Correction to Gaia Absolute Reference Frame}

\begin{figure}
    \ifTHESIS
	    \includegraphics[width=\columnwidth]{Figures_c22a/offset_map_example_ForThesis.pdf}
	\else 
        \includegraphics[width=\columnwidth]{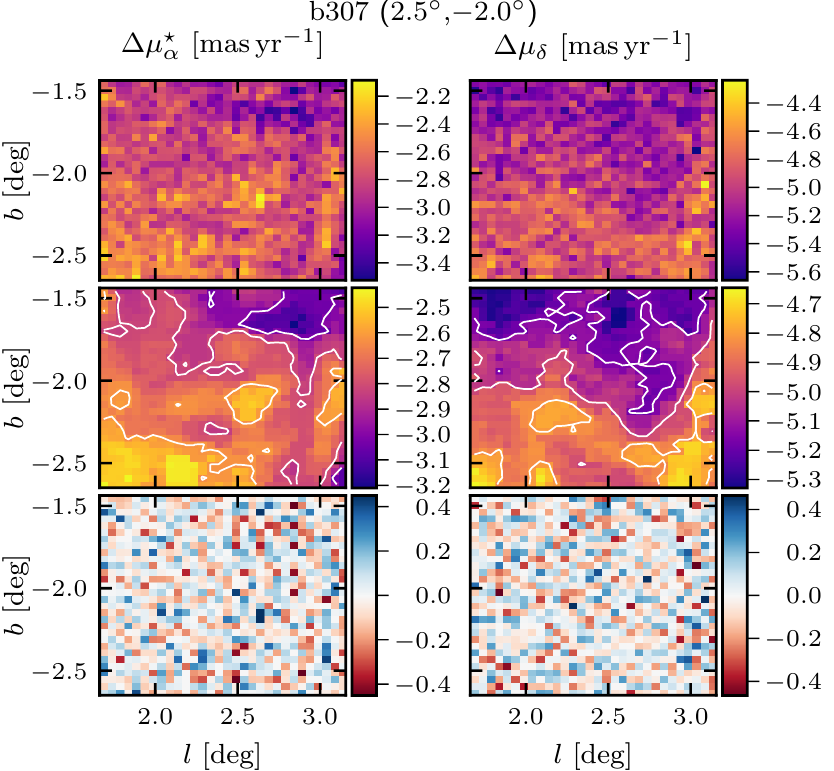}
	\fi 
    \caption[Spatial variation of the correction to the \textit{Gaia} absolute reference frame.]{
    Spatial variation of the correction to the \textit{Gaia} absolute reference frame, for RA (left) and DEC (right), in an example field (b307) on a $30\times30$ sub-tile grid.
    \textit{Top:} Mean offset maps between VIRAC and \textit{Gaia} proper motions ($\Delta\mu_i$).
    \textit{Middle:} Median-smoothed offset maps ($\widetilde{\Delta\mu_i}$).
    \textit{Bottom:} Residual maps showing the stochastic variation of the offset from the median-smoothed maps ($\Delta\mu_i - \widetilde{\Delta\mu_i}$).
    From the smoothed map one can see significant spatial correlations which reduces the number of effective independent regions in the maps.
    }
    \label{c22a:fig:gaia_ref_offset}
\end{figure}

\paragraph{Spatial Variation over a Tile\\}\label{c22a:paragraph:gaia_spatial}
VIRAC relative proper motions are shifted onto the \textit{Gaia} reference frame using a single correction vector per tile.
Were both VIRAC and \textit{Gaia} on perfect, internally consistent, reference frames the computed vector would be constant over a tile.
This is not the case as shown in \cref{c22a:fig:gaia_ref_offset} where we divide the map onto a 30x30 grid.
The top row shows the spatial variation of the correction vector within a single tile. 
There is significant, up to $\sim 1\masyr$, variation which naturally introduces an error into the mean proper motions but the spread in correction vector also adds a broadening effect to the observed proper motion dispersions as some stars are shifted too much, others not enough.

The second row of \cref{c22a:fig:gaia_ref_offset} shows median-smoothed offset maps in which clear, large scale, correlated variations are apparent. 
The bottom row shows the residual between the original and smoothed maps which is the approximately stochastic fluctuation in the offset. 
The presence of spatial correlations is most likely caused by differences in the VIRAC reference frame on different detector chips \citetext{L.C. Smith, private communication}.
These correlations mean we must split the uncertainty into two effects; the stochastic part, with dispersion $\sigma_\mathrm{stat}$, and the spatially correlated part, with dispersion $\sigma_\mathrm{corr}$.

The error on the dispersion is then easily calculated; we define a broadening width, $f_i=\sqrt{\sigma_\mathrm{stat}^2 + \sigma_\mathrm{corr}^2}$, ($i \in \left\{l,\,b\right\}$) which then allows us to estimate the error on the dispersion as described in \cref{c22a:subsubsec:gVIRAC_broad}.
While $f_{i}$ can be as large as $\approx 0.4 \masyr$, the convolution with a velocity distribution with intrinsic dispersion of $3.0 \masyr$ results in an increase in the dispersion of only $\sqrt{ 3^2 + 0.4^2 } - 3 \simeq 0.027 \masyr$ which is relatively small, see \cref{c22a:subsubsec:vvv_err_dists}.

The error on the mean proper motion, $\delta_{\mpml}$ is more complex. We use the standard error on the mean\footnote{
The standard error on the mean, $\mathrm{SE}_{\bar{x}}$, is statistically well defined for a set of $n$ independent measurements, given by $\mathrm{SE}_{\bar{x}} = \sqrt{ \mathrm{var}(x) / n }$. This simple relation fails when the points are no longer independent. 
}
in each case; for the stochastic fluctuation $\sqrt{N}=30$ as the points are independent while for the correlated fluctuations we visually determine the number of \textit{effective} data points to be $N^\star=16$.
We therefore have,
\begin{linenomath}
\begin{equation}
    \delta_{\mpml , \, i} = \sqrt{ \left(\frac{\sigma_{i,\,\mathrm{stat}}}{30}\right)^2 + \left( \frac{\sigma_{i,\,\mathrm{corr}}}{\sqrt{16}} \right)^2}.
\end{equation}\end{linenomath}

\paragraph{Variation with Magnitude\\}\label{c22a:paragraph:gaia_magntiude}

\begin{figure}
    \ifTHESIS
    	\includegraphics[width=\columnwidth]{Figures_c22a/gaia_correction_vectors_Galactic_ForThesis.pdf}
    \else
    	\includegraphics[width=\columnwidth]{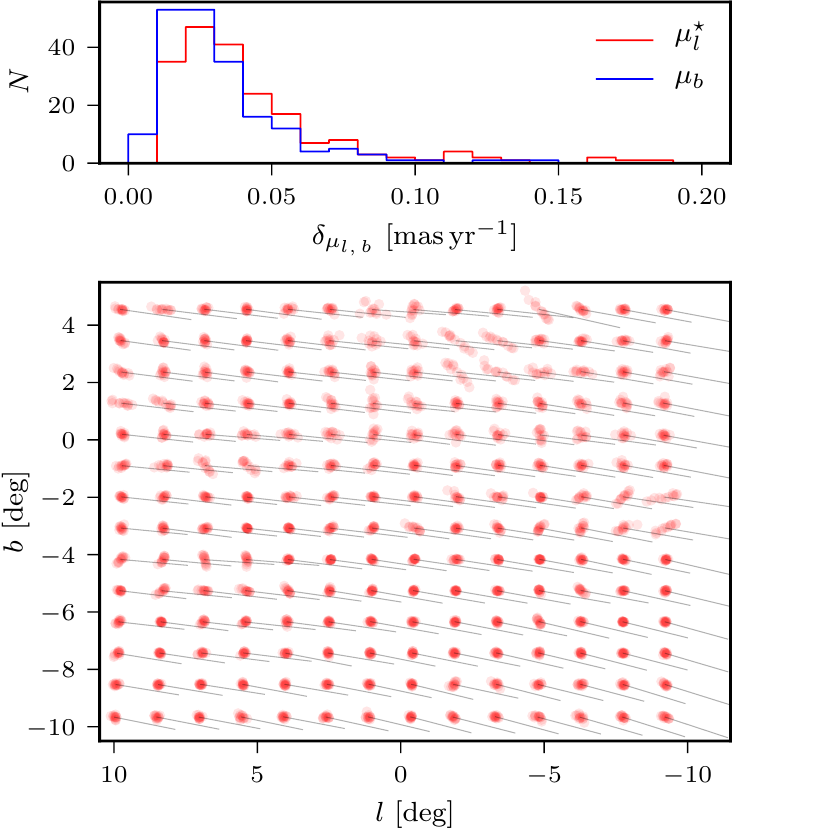}
    \fi
    \caption[Magnitude dependence of the VIRAC to \textit{Gaia} reference frame correction vector.]{Magnitude dependence of the VIRAC to \textit{Gaia} reference frame correction vector.
    \textit{Bottom}: Overall VIRAC to \textit{Gaia} correction vectors are shown as grey lines.
    The red dots, placed at the tile centre for convenience, represent the distribution of alternative endpoints of the vector when it is calculated as a function of magnitude. A minority of tiles have correction vectors that are highly dependent on the magnitude interval used to compute it. Spatially these correspond exactly to clear irregularities in the kinematic maps, for an example see \citepalias[][Fig. 10 top left panel]{clarke_2019}.
    \textit{Top}: Uncertainty distributions for $\mul$ (red), and $\mub$ (blue), due to the magnitude dependence of the correction vector.
    }
    \label{c22a:fig:gaia_ref_offset_fmag}
\end{figure}

In addition, we have found a magnitude-dependent effect in the reference
frame correction.
When correcting to the \textit{Gaia} reference frame we consider stars in the magnitude range $12.5 \leq \ks \,\, \magnB \, \leq 15.0$.
\cref{c22a:fig:gaia_ref_offset_fmag} shows the correction vectors as a function of magnitude.
At high latitude, $|b| \gtrapprox -5\dg$, the correction is approximately magnitude independent. However some tiles closer to the plane exhibit significant variation in the correction vector as a function of magnitude, implying a systematic, magnitude dependent effect in the gVIRAC data. 
The uncertainty distribution for each coordinate axis is shown in the top panel; the uncertainty for each tile is the standard deviation of the magnitude dependent correction vectors, weighted by number of stars in the magnitude interval.
We take this as an estimate of the uncertainty in the overall correction vector.
The median error is $\delta_{\mu_{l,\,b}}\approx 0.03 \masyr$ (the majority of fields do not particularly suffer from this effect) but in a few extreme cases the error can be as large as 0.15 to 0.20 $\masyr$. These errors can be directly applied to the mean proper motion and we apply the \cref{c22a:subsubsec:gVIRAC_broad} approach to determine the dispersion error.

\subsubsection{Differential broadening in RC\&B Extraction}\label{c22a:subsubsec:rcandb_extract}

\begin{figure}
    \ifTHESIS
    	\includegraphics[width=\columnwidth]{Figures_c22a/RCandB_extraction_error_ForThesis.pdf}
    \else
    	\includegraphics[width=\columnwidth]{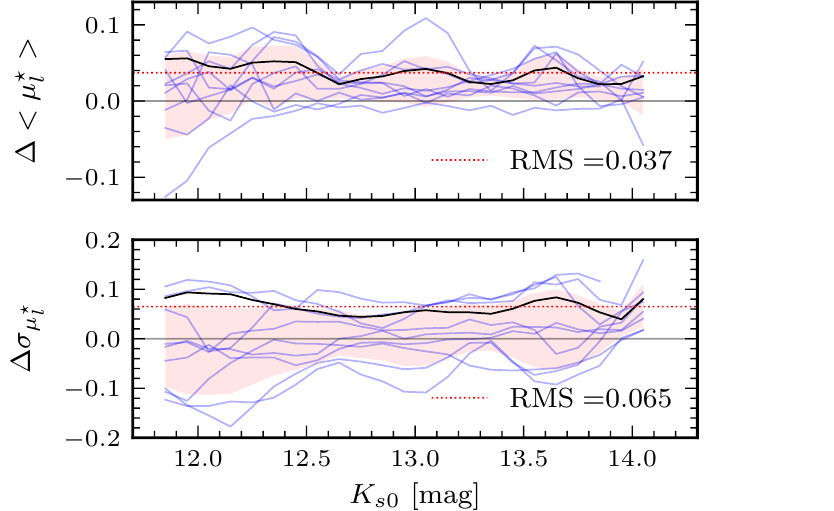}
    \fi
    \caption[Error induced by differential broadening of velocity distribution along LOS.]{
    Errors caused by the RGBC subtraction procedure for obtaining distance-resolved kinematics, using simulations of 9 example fields. 
    \textit{Top:} The difference in mean proper motion between the error convolved RC\&B kinematics and those derived following the approach used on the gVIRAC data.
    \textit{Bottom:} The same as the top panel but for the proper motion dispersion. Blue lines show the profiles of the individual tiles used in the simulation. The shaded pink region outlines $1\sigma$ around the running mean. The solid black line shows the running RMS and the dotted red line shows the magnitude-averaged RMS. We take the RMS values as quoted in the plot as the uncertainty values due to the RGBC subtraction for all tiles (vertical lines in \cref{c22a:fig:vvv_err_dists}).
    }
    \label{c22a:fig:RCandB_extrac_err}
\end{figure}

From the absolute proper motions, RC\&B distance-resolved kinematics are determined as in \citetalias[][(their Section 5.2)]{clarke_2019}, see also \cref{c22a:subsec:gVIRAC}.
As measurement uncertainties generally increase with apparent magnitude the RGBC velocity distribution is broadened to a greater extent at faint magnitudes than at brighter magnitudes. This differential broadening introduces a systematic error into the RC\&B kinematic measurements.

To understand this effect, and to estimate the errors introduced by it, we simulate it using the M2M model. Our approach is as follows:
\begin{inparaenum}[(i)]
    \item sample particles from the model for nine representative LOS; using the different stellar type synth-LFs (see \cref{c22a:tab:acronyms_c22a}) we can construct the overall RGB and RC\&B proper motion distributions at each magnitude;
    \item broaden these distributions by taking the median proper motion uncertainty of the corresponding gVIRAC data, $\varepsilon\left(l,\,b,\,\ks\right)$, and adding a random shift, $\Delta \mu_i \sim \mathcal{N}\left(0,\varepsilon\left(l,\,b,\,\ks\right)\right)$, to each proper motion;
    \item compare the mean and dispersion of the error-convolved RC\&B distributions to the values obtained by applying the RGBC-subtraction method \citepalias{clarke_2019} to the simulated RGB proper motion distributions.
\end{inparaenum}
The difference in the mean (dispersion) is shown in the top (bottom) panel of \cref{c22a:fig:RCandB_extrac_err}.
There is an average positive shift in $\mpml$ while the dispersions exhibit no obvious structure.
We therefore use the magnitude integrated RMS, see \cref{c22a:fig:RCandB_extrac_err}, as a constant error factor for all 196 LOS.
This approach smooths out the fluctuations in the simulated error which are likely caused by the limited number of particles in the M2M model.
The uncertainty on the mean (dispersion) is $\delta_{<\mu_i>} = 0.037\masyr$ ($\delta_{\sigma_i}=0.065\masyr$).

\subsubsection{Statistical Errors on RC\&B Kinematic Measurements}\label{c22a:subsubsec:virac_stat_err}

By kernel-smoothing the RGBC velocity distribution (at faint magnitudes), and subtracting it from the smoothed RGB, \citetalias{clarke_2019} obtained the kernel-smoothed RC\&B velocity distribution.
The RC\&B mean and velocity dispersion were then computed by numerical Monte Carlo re-sampling of the smoothed RC\&B velocity distribution, see \citepalias[][Section 5.2]{clarke_2019} for further details.
To avoid constructing a re-sampled velocity distribution that is too well characterised or vice versa, the number of RC\&B stars that are re-sampled is set equal to the number of excess stars above the exponential fit to the RGBC. Repeated re-samplings allows us to define the mean, dispersion, and suitable errors.
This approach builds in a dependence on the $\frcb$ as, for a given number of stars, a voxel with a larger $\frcb$ will have a better defined RC\&B velocity distribution and thus smaller errors on the mean and dispersion.

\subsubsection{gVIRAC Combined Error Distributions}\label{c22a:subsubsec:vvv_err_dists}

\begin{figure*}
    \ifTHESIS
	    \includegraphics[width=\textwidth]{Figures_c22a/VVV_err_dist_FuncMag_RCandB_ForThesis.pdf}
	\else
        \includegraphics[width=\textwidth]{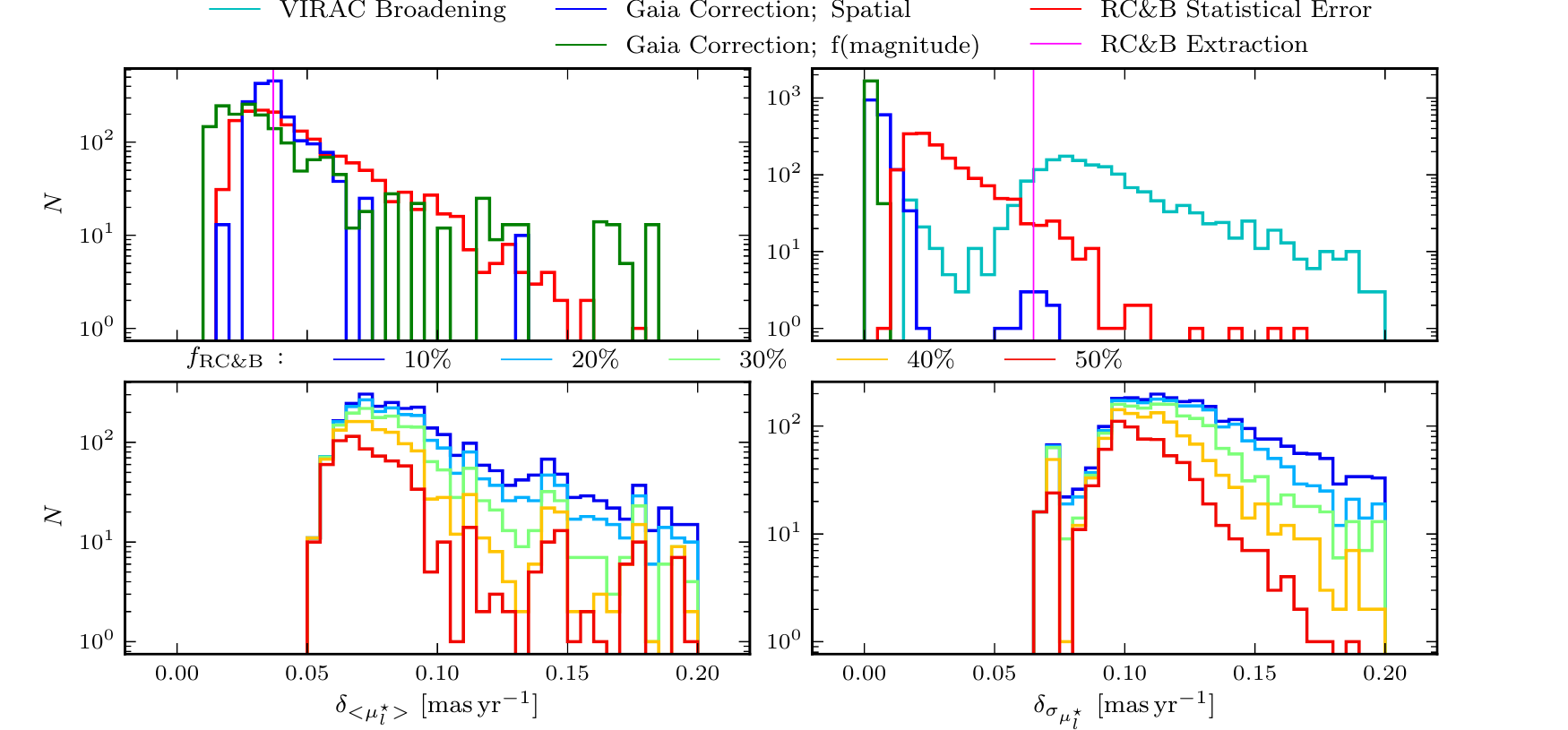}
    \fi
    \caption[Uncertainty distributions for the gVIRAC data.]{
    Distributions of uncertainties for the gVIRAC data across all voxels $(l,b,K_{s0})_i$: $\mpml$ (left) and $\dpml$ (right).
    \textit{Top:} 
    Uncertainties from individual sources for the $\frcb=30\%$ case. In the case of RC\&B extraction we plot a vertical line at the single value we adopt and use for all tiles. VIRAC broadening does not affect the mean proper motion and so does not contribute in the left column.
    \textit{Bottom:} Total uncertainty, derived by addition in quadrature, for each of the five $\frcb$ masks considered in this work. Stricter cuts restrict the inclusion of high error voxels to a greater extent when compared to lower error voxels.
    }
    \label{c22a:fig:vvv_err_dists}
\end{figure*}

Histograms of the different error contributions for $\mpml$ (left column) and $\dpml$ (right column) are shown in the top row of \cref{c22a:fig:vvv_err_dists}. The bottom row shows the total error (via summation in quadrature) for different $\frcb$ masks. 
The median uncertainties for the $\frcb=30\%$ case are $\delta_{\mpml}\!\approx\!0.070\!\masyr$ and $\delta_{\dpml}\!\approx\!0.105\!\masyr$.
The total $\mpml$ error is an approximately balanced combination of the four sources with each contributing roughly equally around the $\approx \!0.03\!\masyr$ level. The $\dpml$ uncertainty is dominated by:
\begin{inparaenum}[(i)]
    \item the broadening by individual proper motion uncertainties, and
    \item the RGBC subtraction uncertainty,
\end{inparaenum}
which both contribute at $\gtrapprox 0.07 \masyr$.

Our distance resolved kinematics consider RC\&B stars; voxels in which we have a large $\frcb$ have, in general, better determined kinematic measurements.
The $\mpml$ error is generally $0.05 \lessapprox \delta_{\mpml} \,\,\masyrB \lessapprox 0.10$ however for smaller $\frcb$ there is a substantial tail to high error. The dispersion error is similar; generally $0.09 \lessapprox \delta_{\dpml} \,\,\masyrB \lessapprox 0.17$ but with a large tail to high error. In both cases using a stricter $\frcb$ criteria shifts the median error of the distribution to smaller values; unsurprising given the $\frcb$ criteria defines where the RC\&B kinematics are best known. 
Specifically, the statistical measurement uncertainties depend on $\frcb$ as voxels with a relatively smaller $\frcb$, for a given total number of stars, have fewer RC\&B stars with which to measure the mean and dispersion. 
As discussed in \cref{c22a:subsec:RCBfrac} we see that using small $\frcb$ fractions permits a disproportionate number of high error voxels relative to the stricter cases. This is especially true for the dispersions.

\subsection{Sources of Error in the Models}\label{c22a:subsec-modelerr}

\subsubsection{Luminosity Function \& Bar Angle} \label{c22a:subsubsec:lf_bangle_error}

\begin{figure}
    \ifTHESIS
        \includegraphics[width=\columnwidth]{Figures_c22a/CompareShifted_LFs_4paper_ForThesis.pdf}
    \else
    	\includegraphics[width=\columnwidth]{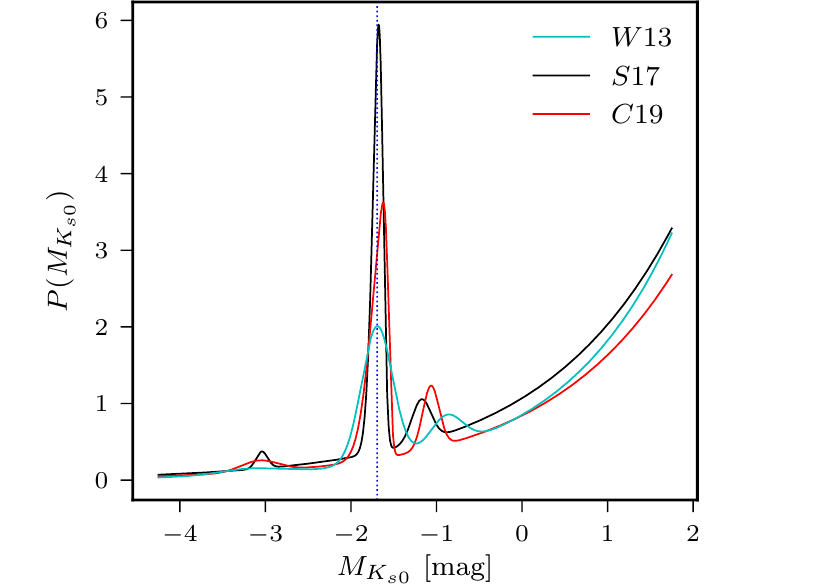}
    \fi
    \caption[Comparison between three plausible luminosity functions.]{
    Comparison between the three synth-LFs considered in this analysis. In the legend S17 refers to the synth-LF of \citetalias{simion_2017} while the other labels are as defined in the text. Each synth-LF is shifted such that $<M_{K_{s0},\,\mathrm{RC}}>=-1.694$ mag (vertical dashed blue line). We use the \citetalias{wegg_2013} synth-LF when computing fiducial model predictions.
    }
    \label{c22a:fig:lum_funcs}
\end{figure}

We make two assumptions when predicting kinematics from the M2M models; the choice of synth-LF and the bar angle, $\bangle$. 
We take the (\citetalias{wegg_2013} synth-LF, $\bangle=28^\circ$) combination as our fiducial assumption as the \citetalias{portail_2017a} models are fit to a density distribution produced by de-convolving the VVV obs-LFs with the \citetalias{wegg_2013} synth-LF. By re-convolving using the same synth-LF, we will recover the true VVV obs-LF.

\cref{c22a:fig:lum_funcs} shows three recent examples of synth-LFs generated for the MW bulge region using slightly different assumptions on the metallicity distribution and the choice of stellar isochrones. 
There are clear differences: 
\begin{inparaenum}
    \item the width of the RC;
    \item the magnitude of the RGBB relative to the RC;
    \item the strength of the AGBB;
    \item the shape of the RC; and
    \item the shape of the RGBC.
\end{inparaenum}
The choice of synth-LF impacts the predicted kinematics, for example a wider RC component allows a particle to contribute to the kinematics at a larger range of apparent magnitudes than a thinner component. As we use the RC\&B, not just the RC, the RGBB and AGBB must also be considered.

The choice of $\bangle$ affects both the observed kinematics and the observable LOS density distribution. Observing a bar at a more end-on angle projects less of the bar streaming velocity into proper motion (the radial velocity increases).
An edge-on bar, $\bangle=90\dg$, exhibits the narrowest LOS density distribution because the LOS is approximately perpendicular to the bar major axis. 
Changing the synth-LF, with no corresponding change to $\bangle$, changes the width of the obs-LF. 
However, using a synth-LF with a narrower RC can approximately compensate for the differences induced by using a smaller $\bangle$ value (more end-on).

We therefore consider three basic combinations of synth-LF and $\bangle$; 
\begin{inparaenum}
    \item the \citetalias{wegg_2013} synth-LF with $\bangle=28\dg$,
    \item the \citet[hereafter S17]{simion_2017} synth-LF with $\bangle=22\dg$ as was found to be best by \citet{sanders_2019a}, and
    \item the \citetalias{clarke_2019} synth-LF with, given the synth-LF is intermediate between those of \citetalias{simion_2017} and \citetalias{wegg_2013}, the intermediate $\bangle=25\dg$.
\end{inparaenum}

To derive the uncertainty introduced by the  of synth-LF and $\bangle$ we consider all three synth-LFs and additionally vary the $\bangle$ value by $\pm2^\circ$ around the optimum. This results in nine predictions of the mean proper motion and dispersion for each voxel.

The error due to the bar angle is determined by first taking the standard deviation over bar angles in each voxel, resulting in three $\delta_{\bangle}$ values corresponding to each of the three synth-LFs. Taking the mean of these three values gives the error introduced by the choice of $\bangle$ marginalised over synth-LF.

The error introduced by the choice of synth-LF is determined in similar fashion. We first take the mean over bar angles in each voxel, obtaining predictions for each synth-LF marginalised over $\bangle$, and then take the standard deviation of the three values to obtain $\delta_\mathrm{synth-LF}$ for each voxel.

\subsubsection{M2M Modelling Errors}\label{c22a:subsubsec:m2m_errors}

The M2M method used by \citetalias{portail_2017a} works by gradually adjusting particle weights such that the $\chi^2$ between data observables and model predictions is minimised.
There is an intrinsic error in the model predictions due to the non-perfect convergence of the particle weights to final values; the particle weights oscillate slightly around their long term values.
This oscillation translates to a snapshot to snapshot fluctuation in model predictions.
Once the model has stabilised and the particle weights are fluctuating around their long term values there remains a uncertainty due to how long one continues to apply the model fitting. 
Numerical effects, and gradual changes to the dynamical structure of the model, can both affect the predicted kinematics.
We account for these effects by comparing the predictions of a single model, $\omegab=37.5\kmskpc$, and $\vphisun=247.5\kms$, at 21 snapshots separated by 500 fitting iterations. The separation between each snapshot corresponds to $\approx 0.85\tau_\mathrm{dyn}$ (dynamical times\footnote{Dynamical time is determined using $\tau_\mathrm{dyn} = 2\pi R / \vcirc \approx 65 \myr$ with $R=2\kpc$ and $\vcirc(R=2\kpc)=190\kms$ \citepalias[][fig.23]{portail_2017a}.}) and the total period corresponds to $\approx 17\tau_\mathrm{dyn}$. The voxel-wise error is the standard deviation of all predictions for each voxel. This approach simultaneously captures the stochastic fluctuation of the model predictions due to the non-perfect convergence of the particle weights and the systematic shift of the model predictions due to long term changes to the model structure.

The final stage in a M2M fit evolves the model for a short time without fitting; the particles phase-mix to a final steady state, often a slightly worse fit than when fitting, during which the model predictions change.
To account for the change in the model predictions we compare eight snapshots taken during the phase-mixing step, each separated by 1000 iterations. The corresponding voxel-wise uncertainty in the model predictions is the standard deviation of the model predictions.

\subsubsection{Combined Model Error Distributions}\label{c22a:subsubsec:mod_err_dists}

\begin{figure*}
    \ifTHESIS
	    \includegraphics[width=\textwidth]{Figures_c22a/MODEL_err_dist_FuncMag_RCandB_ForThesis.pdf}
	\else
	    \includegraphics[width=\textwidth]{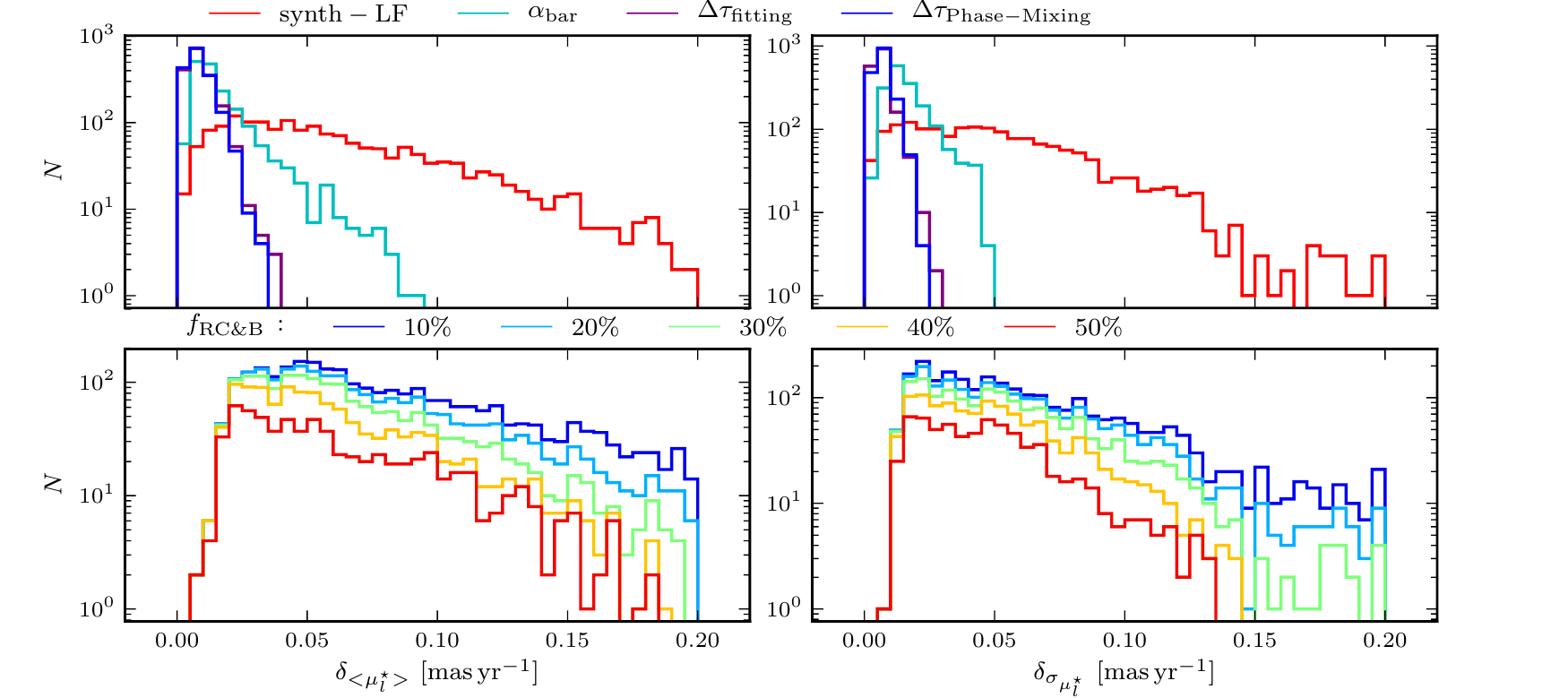}
	\fi
    \caption[Uncertainty distributions for model related effects.]{
    Same as \cref{c22a:fig:vvv_err_dists} but for the model errors.
    }
    \label{c22a:fig:mod_err_dists}
\end{figure*}

The model-based error distributions are shown in \cref{c22a:fig:mod_err_dists}. For both $\mpml$ and $\dpml$ the dominant source of error is the choice of synth-LF followed by the choice of $\bangle$. 
This is expected as the synth-LF, despite all being realistic possibilities, are distinct while the choice of $\bangle$ produces a more gradual change in predicted kinematics. The choice of synth-LF and $\bangle$ produces errors generally larger than the modelling errors as, with $10^6$ stellar particles, the models are well defined and relatively stable.
The $\frcb=30\%$ median errors are $\delta_{\mpml}\approx 0.06\masyr$, and $\delta_{\dpml}\approx 0.05\masyr$. The phase-mixing and fitting-length errors generally contribute in the region $0.00\lesssim \delta \,\,\masyrB \lesssim 0.02$, the $\bangle$ error only slightly larger than that, albeit with a larger high-error tail. Despite using appropriate $\bangle$ values for each synth-LF, the choice of synth-LF dominates the error.

The total error distributions for different $\frcb$ are shown in the bottom row of \cref{c22a:fig:mod_err_dists}. The long tails observed for the less strict, up to $\frcb>30\%$, criteria are caused by the error in the choice of synth-LF. For both $\mpml$ and $\dpml$ the median overall error is smaller than the corresponding data-associated errors which is encouraging.

\section{Model-Data Comparison}\label{c22a:sec:statistics}

We compare the M2M models with the data using the mean proper motion $\mpml$ and dispersion $\dpml$ of the RC\&B population across the VIRAC tiles, in \textit{voxels} $(l, b, K_{s0})_i$ (\cref{c22a:subsec:gVIRAC}). The model dispersions are convolved with the respective median VIRAC proper motion errors (\cref{c22a:subsubsec:gVIRAC_broad}). All error contributions from \cref{c22a:sec:errors}, both data based and model based, are combined into a single uncertainty for each voxel, adding them in quadrature. We adopt an outlier-tolerant likelihood approach which allows for possible additional systematic errors by treating the voxel uncertainties as lower bounds on their true values \citep{sivia_2006}.

\subsection{An Outlier-Tolerant Approach} \label{c22a:subsec:outliertol}

Here we present in more detail the statistical framework used for the quantitative comparison of the P17 models with the gVIRAC data.
As illustrated in \citetalias{clarke_2019}, and shown more quantitatively in \cref{c22a:subsec:fiducial}, the models fit the gVIRAC data well despite not being fit to the data. However there do remain some regions with high residuals (see \cref{c22a:fig:voxelwise_logL_map} and \cref{c22a:subsec:fiducial}).
These remaining large residuals result in large $\chi^2$ values which, if unaccounted for, could bias the final result.
To overcome this we apply an outlier-tolerant likelihood-based approach described as a \textit{conservative formulation} by \citet{sivia_2006} and applied, e.g., by \citet{reid_2014} to model masers in Galactic spiral arms. The uncertainties in each voxel are treated as a lower bound on the true uncertainty. The likelihood function (which must be maximised) for the $i^\mathrm{th}$ voxel is given by \citep{sivia_2006},
\begin{linenomath}\begin{equation}
    \lhood_i\left(d_i|\boldsymbol{\theta}, \delta_i\right) = \frac{1}{\sqrt{2\pi}\delta_i}
    \left[
    \frac{ 1 - e^{-\chi_i^2 / 2} }{\chi_i^2}
    \right],
\end{equation}\end{linenomath}
where,
\begin{linenomath}\begin{equation}\label{c22a:eqn:chi_i}
    \chi_{_i} = \frac{ d_i - m_i\left(\boldsymbol{\theta}\right) }{\delta_i},
\end{equation}\end{linenomath}
$d_i$, $\delta_i$, $m_i(\boldsymbol{\theta})$ are the $i^{\rm th}$ values of the data $d$, error $\delta$, and model $m$. Here $\delta_i = \sqrt{{\delta_{d,i}}^2 + {\delta_{m,i}}^2}$ is the combined data and model error\footnote{Note that we use $\delta$, rather than $\sigma$, to represent errors in mean and dispersion to avoid confusion as $\sigma$ denotes the intrinsic dispersion of a proper motion distribution.}, and $m_i(\boldsymbol{\theta})$ is the prediction of the model given model parameters, $\boldsymbol{\theta} \equiv (\omegab,\,\,\vphisun)$.

The overall log-likelihood is then given by,
\begin{linenomath}\begin{equation}
    \log_\mathrm{e}\left[\,\,
    \lhood\left(
    \{d\}|\boldsymbol{\theta},\{\delta\}
    \right)\,\,
    \right]
    = 
    \sum_{i=1}^{N} 
    \log_\mathrm{e} \left(
    \frac{ 1 - e^{-\chi_i^2 / 2} }{\chi_i^2 }\frac{1}{\sqrt{2\pi\delta_i^2}}
    \right).
\end{equation}\end{linenomath}
From Bayes theorem,
\begin{equation}
    P\left(m|d\right) = \frac{P\left(d|m\right)P\left(m\right)}{P\left(d\right)},
\end{equation}
the posterior probability is
\begin{equation}
    \log_e\left[ P\left( \boldsymbol{\theta} | \left\{d,\delta\right\} \right)\right] 
    \propto
    \log_\mathrm{e}\left[
    \lhood\left(
    \{d\}|\boldsymbol{\theta},\{\delta\}
    \right)
    \right]
    + 
    \log_e\left[  
    \pi\left(
    \boldsymbol{\theta}
    \right)
    \right],
\end{equation}
where we drop the normalising $P(d)$ evidence term and $\pi ( \boldsymbol{\theta})$ denotes any prior on $\omegab$ and $\vphisun$. Our fiducial assumption is to adopt uninformative priors, $\pi(\boldsymbol{\theta}) \sim \fU_{\omegab}(30.0,45.0) \cdot \fU_{\vphisun}(240.0,260.0) $, however we also investigate the effect of $\pi(\vphisun) \sim \fN(250.63,0.42)$, the constraint on $\vphisun$ from \citetalias{gravity_2020} and \citetalias{reid_2020}, and of $\pi(\omegab|\vcirc)$, the probability of the different model rotation curves using the data from \citet{eilers_2019} and \citet{reid_2019}.

To locate the maximum-posterior point in parameter space and determine confidence intervals we require higher resolution than provided by the grid of models. To remedy this we interpolate between the models onto a high-resolution grid.
Interpolation is plausible in this case as, due to the models' construction, the $\logl$ varies smoothly over ($\omegab$, $\vphisun$) parameter space.
We obtain constraints on $\omegab$ ($\vphisun$) by marginalising over $\vphisun$ ($\omegab$), normalising the posterior probability curve so that the total area integrates to unity, and then locating the narrowest region in parameter space in which the area integrates to $\mathrm{erf}(1/\sqrt{2})\approx0.683$.

\subsection{Fiducial Case}\label{c22a:subsec:fiducial}

\begin{figure}
    \ifTHESIS
	    \includegraphics[width=\columnwidth]{Figures_c22a/log_L_curves_W13_ForThesis.pdf}
    \else
	    \includegraphics[width=\columnwidth]{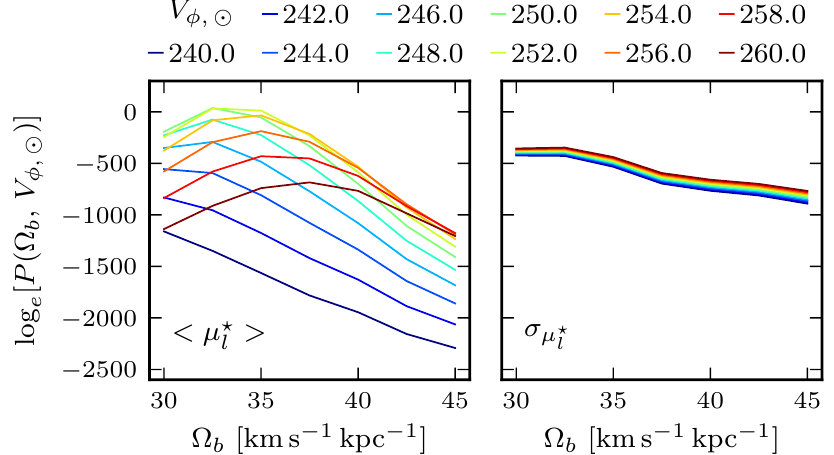}
	\fi
    \caption[Posterior probability curves for {$\vphisun$} \& {$\omegab$} independently.]{
    The posterior probability curves for the best model under fiducial assumptions. The $\mpml$ data (left) provides the majority of the constraining power and has a clearly defined maximum. The $\dpml$ data (right) is less constraining and the posterior for the best model is significantly more negative than for the $\mpml$ data. There is also no clearly defined maximum with the dispersion preferring larger values of $\vphisun$. See \cref{c22a:subsec:fiducial} for discussion.
    }
    \label{c22a:fig:logL_curves_fiducial}
\end{figure}

Here we present the results for the fiducial comparison of the \citetalias{portail_2017a} models with the gVIRAC data. The underlying assumptions, varied and tested in \cref{c22a:sec:testing} below, are:
\begin{inparaenum}
    \item only voxels are included in which $f_\mathrm{RC\&B} > 30\%$;
    \item the \citetalias{wegg_2013} synth-LF is used in the models, see \cref{c22a:subsubsec:lf_bangle_error}, together with
    \item the corresponding bar angle $\bangle=28\dg$ (\citetalias{portail_2017a}).
\end{inparaenum}
\cref{c22a:fig:logL_curves_fiducial} shows the posterior curves for the best model obtained with these assumptions. It is clear that the majority of the gVIRAC constraining power comes from $\mpml$, with $\dpml$ having no clear maximum, preferring slightly smaller $\omegab$ values, at lower maximum posterior probability. The underlying cause is that the model $\dpml$ are systematically slightly too high outside the bulge. While the effect is not large, with typical $\dpml$ errors $<5\%$ it can have some impact. Therefore in the fiducial case we (iv) consider only $\mpml$, and then treat the difference caused by including, or not, the $\dpml$ data as an additional uncertainty. The shift in the measured values induced by including the $\dpml$ data is  $\Delta\omegab=-\psdeltaDisp\kmskpc$ and $\Delta\vphisun=-\vtdeltaDisp\kmskpc$.

\begin{figure}
    \ifTHESIS
	    \includegraphics[width=\columnwidth]{Figures_c22a/log_L_Map_W13_ForThesis.pdf}
    \else
	    \includegraphics[width=\columnwidth]{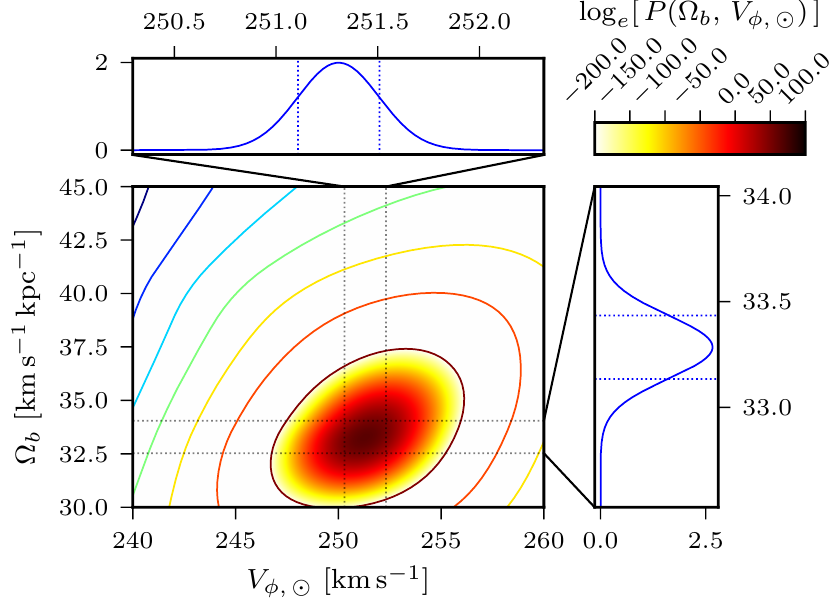}
	\fi
    \caption[Posterior probability distribution in {$\vphisun$}, {$\omegab$} space with marginalised confidence intervals.]{
    Map of $\boldsymbol{ \log{_e} \left[ P\left( \omegab,\, \vphisun \right) \right] }$ for the grid of models. We marginalise over each dimension in turn to locate the $1\sigma$ confidence interval. These intervals are shown in the panels at the top and to the right where we show a zoom-in of the relevant axis (denoted on the 2d map by the dotted lines). We locate the shortest interval containing a total probability of $\approx0.68$ and this is shown by the vertical dotted lines in the zoom panels.
    The region around the maximum-posterior is shaded according to the colourbar while the remainder of the surface is shown by the contours.
    }
    \label{c22a:fig:logL_map_fiducial}
\end{figure}

\cref{c22a:fig:logL_map_fiducial} shows the $\boldsymbol{\logposterior}$ map computed using the outlier-tolerant approach.
This map is \textit{not} normalised however the additional panels show the marginalised, normalised posterior distributions for $\vphisun$ (top) and $\omegab$ (right).
The region around the maximum-posterior is highlighted by the shaded region while the rest of the $\boldsymbol{\logposterior}$ surface is shown by the contours.
The extent of the marginalised panels is shown by the dashed lines on the map.
The normalisation sets the integral under each curve to unity; this is a safe assumption because the posterior probability becomes rapidly negligible away from the maximum, as can be seen in the marginalised panels.
The results we obtain are $\omegab= \psfiducial \kmskpc$, and $\vphisun= \vtfiducial \kms$, see the top row of \cref{c22a:tab:results}.

\begin{figure*}
    \centering
    \ifTHESIS
    	\subfigure{\includegraphics[width=\textwidth]{Figures_c22a/voxelwise_ML_residual_map_W13_ForThesis.pdf}}
        \subfigure{\includegraphics[width=\textwidth]{Figures_c22a/voxelwise_ML_log_L_map_W13_ForThesis.pdf}}
    \else
    	\subfigure{\includegraphics[width=\textwidth]{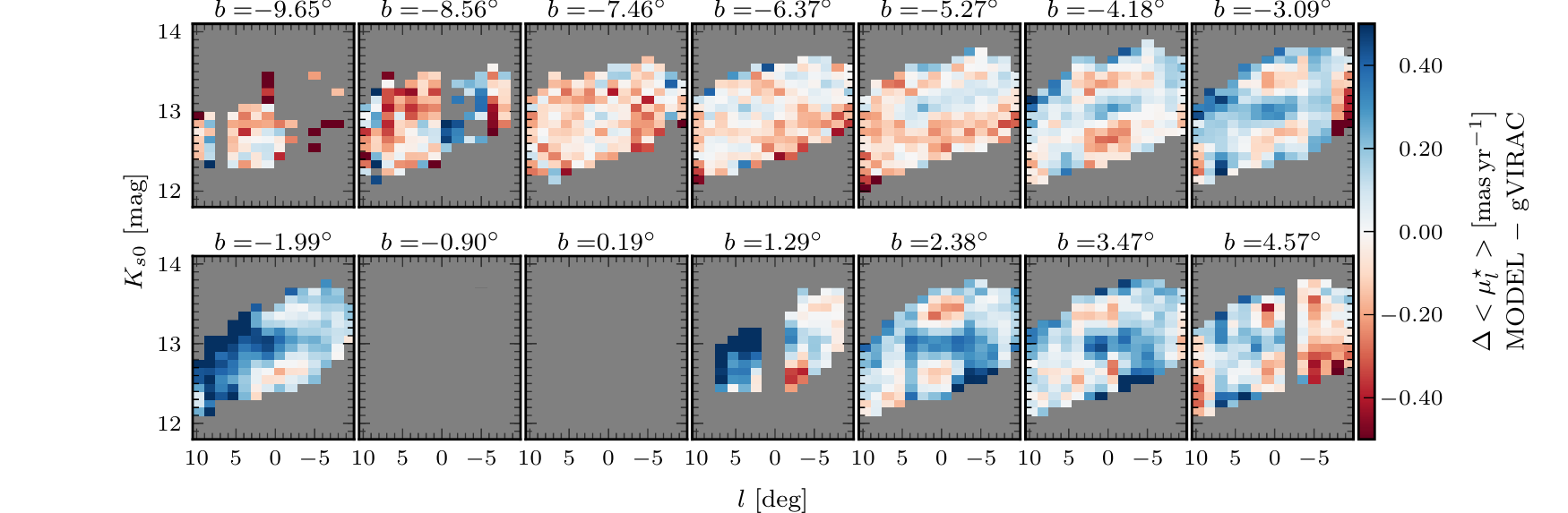}}
        \subfigure{\includegraphics[width=\textwidth]{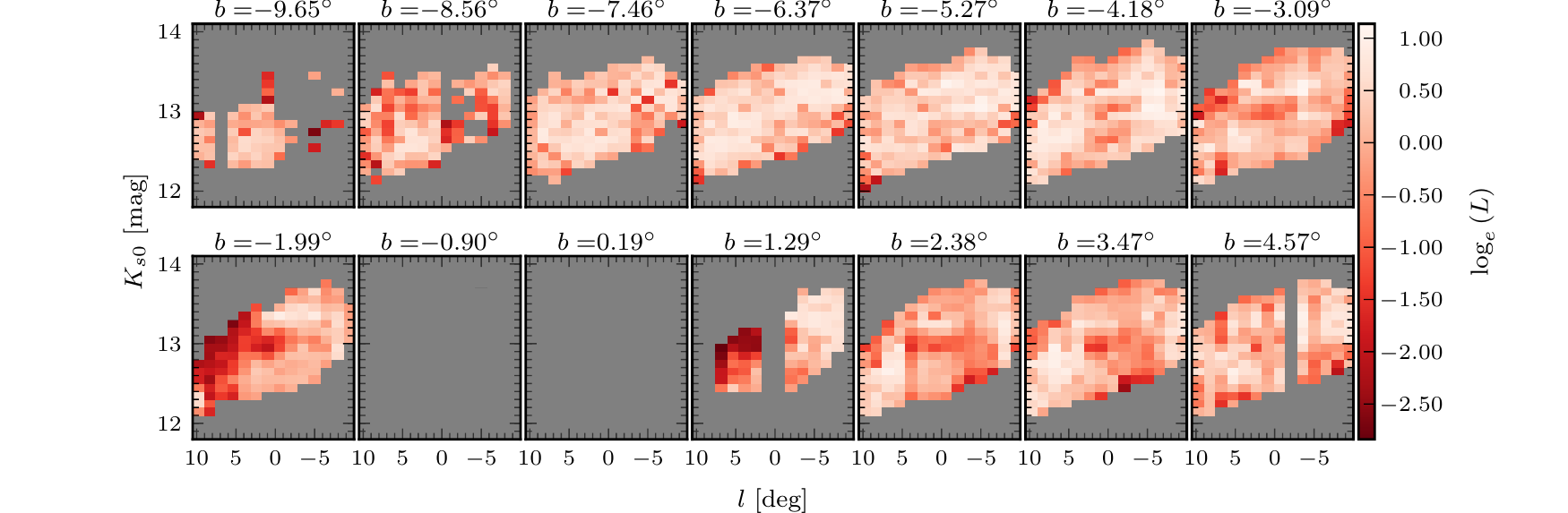}}
    \fi
    \caption[Residual and $\logl$ maps for the best model-data comparison.]{
    \textit{Top}: $\mpml$ residuals (VIRAC-model) for the fiducial model. In general we see excellent agreement between the model and the data; the residuals, when converted to velocity assuming RC star absolute magnitudes, have mean and dispersion $\mu_\Delta=1.2$ \& $\sigma_\Delta=8.9$ $\!\kms$.
    \textit{Bottom}: voxelwise map of the $\logl$ in the fiducial case. 
    To aid conversion to standard $\chi^2$; for a reasonable error value, $\sigma_i=0.1$, and a well fit $\chi^2$ value, $=1.2$, we find $\logl\approx0.35$.
    Over many voxels the likelihood is very good, however there are still regions with remaining systematic differences between model and gVIRAC data.
    These could be, for example, due to the effects of possible overlap with spiral structure and systematics in the RC\&B synth-LF, as discussed in \cref{c22a:sec:testing}.
    }
    \label{c22a:fig:voxelwise_logL_map}
\end{figure*}

We show the residuals between the gVIRAC data and the best fitting model in the top panel of \cref{c22a:fig:voxelwise_logL_map}. Over a large range of $l$ and $b$ the model fits very well; converting the residual to $\!\!\kms$ (taking the central apparent magnitude of each bin and converting to distance assuming RC absolute magnitude) we find the residuals have mean and dispersion, $\mu_\Delta=1.2$ \& $\sigma_\Delta=8.9$ $\!\kms$ (the distribution has stronger wings than Gaussian), indicating excellent general agreement between the model and the VIRAC data.

The bottom panel of \cref{c22a:fig:voxelwise_logL_map} shows a map of the $\logl$. 
The model deviates from the gVIRAC data
\begin{inparaenum}
    \item at faint magnitudes, $+l$, near the Galactic plane; and
    \item towards the bright magnitudes at $-l$, seemingly for all latitudes.
\end{inparaenum} 
These remaining differences reflect the inherent systematic differences between the models and the gVIRAC data. As stated the models have not been fit to gVIRAC so some deviation is expected.
In \cref{c22a:sec:testing} we shall analyse the effect of the various assumptions we have made for the fiducial case.

\ifTHESIS
    \begin{landscape}
\fi
\begin{table*}
    \ifTHESIS
        \small
    \fi
 \caption[Results considering various systematic effects and varying fiducial assumptions.]{Results for the pattern speed and azimuthal solar motion derived for the fiducial assumptions and the subsequent variations.}
 \label{c22a:tab:results}
 \begin{tabular}{l@{\hskip 15pt} cccc @{\hskip 15pt} cc}
  \hline\\[-5pt]
    & Data & Mask & (synth-LF, $\bangle$) & Prior &    $\omegab$  & $\vphisun$ \\[2pt]
    &        &      &              $\degB$  &       &    $\kmskpcB$ & $\kmsB$    \\[2pt]
  \hline\\[-5pt] 
   Fiducial & $\mpml$ & $f_\mathrm{RC\&B}=30\%$ & (\citetalias{wegg_2013}, $28$) && $\psfiducial$ & $\vtfiducial$ \\[10pt]
   %
   %
   Fiducial & $\mpml$ \& $\dpml$ & $f_\mathrm{RC\&B}=30\%$ & (\citetalias{wegg_2013}, $28$) && $\psWithDisp$ & $\vtWithDisp$ \\[10pt]
   %
   %
   %
   Vary $f_\mathrm{RC\&B}$ & $\mpml$ & $f_\mathrm{RC\&B}=10\%$ & (\citetalias{wegg_2013}, $28$) && $\psRCandBten$    & $\vtRCandBten$    \\[2pt]
                           &                  & $f_\mathrm{RC\&B}=20\%$ &                                && $\psRCandBtwenty$ & $\vtRCandBtwenty$ \\[2pt]
                           &                  & $f_\mathrm{RC\&B}=40\%$ &                                && $\psRCandBforty$  & $\vtRCandBforty$  \\[2pt]
                           &                  & $f_\mathrm{RC\&B}=50\%$ &                                && $\psRCandBfifty$  & $\vtRCandBfifty$  \\[10pt]
    %
    %
    %
    Partial Data &  $\mpml$($+l$,$\pm b$) \textcolor{blue}{$^\dagger$} &  $f_\mathrm{RC\&B}=30\%$  & (\citetalias{wegg_2013}, $28$) &&  $\psleft$  & $\vtleft$  \\[2pt]
                 &  $\mpml$($-l$,$\pm b$) \textcolor{white}{$^\dagger$}&                           &                                &&  $\psright$ & $\vtright$ \\[10pt]
    %
    %
    %
    Vary LF \& $\bangle$ \textcolor{blue}{$^\star$} & $\mpml$ & $f_\mathrm{RC\&B}=30\%$ & (\citetalias{simion_2017}, $22$) && $\psLFalphaS$  & $\vtLFalphaS$  \\[2pt]
                                                    &                  &                         & (\citetalias{clarke_2019}, $25$) && $\psLFalphaCa$ & $\vtLFalphaCa$ \\[2pt]
                                                    &                  &                         & (\citetalias{clarke_2019}, $28$) && $\psLFalphaCb$ & $\vtLFalphaCb$ \\[10pt]
    %
    %
    %
    Spiral Structure & $\mpml$ & Mask-W15 (Ellipse) & (\citetalias{wegg_2013}, $28$) && $\psWellipseSpiral$ & $\vtWellipseSpiral$ \\[2pt]
                     &                  & Mask-P20 (Contour) &                                && $\psPcontourSpiral$ & $\vtPcontourSpiral$ \\[10pt]
    %
    %
    %
    Prior on $\vphisun$ & $\mpml$ & $f_\mathrm{RC\&B}=30\%$ & (\citetalias{wegg_2013}, $28$) & $\fN(\gravReidVphiVal,\,\gravReidVphiErr)$ & $\psVphiPrior $ & $\vtVphiPrior$ \\[10pt]
    %
    %
    %
    Prior on $\vcirc$\textcolor{blue}{$^\ddag$} & $\mpml$ & $f_\mathrm{RC\&B}=30\%$ & (\citetalias{wegg_2013}, $28$) & $\pi(\omegab|\vcirc)$ E19 & $\psVcircEilersPrior $ & $\vtVcircEilersPrior $ \\[2pt]
                      &                  &                         &                                & $\pi(\omegab|\vcirc)$ R19 & $\psVcircReidPrior $ & $\vtVcircReidPrior $ \\[2pt]
  \hline
 \end{tabular}
 \begin{tablenotes}
      \small
      \item \textcolor{blue}{$^\star$} Here we only quote the values for the $f_\mathrm{RC\&B}=30\%$ case. The results, using the other possible masks, are shown in \cref{c22a:fig:vary_LF_alpha}.
      \item \textcolor{blue}{$^\dagger$} We exclude the 4 most in-plane latitude slices; see text of \cref{c22a:subsec:many_min}.
      \item \textcolor{blue}{$^\ddag$} Here E19 refers to the rotation curve of \citet{eilers_2019} and R19 refers to that of \citet{reid_2019}.
    \end{tablenotes}
\end{table*}
\ifTHESIS
    \end{landscape}
\fi

\subsection{Effect of Priors}\label{c22a:subsec:priors}

One might wonder whether, given the precise measurements of $\ro$ \citepalias{gravity_2020} and the proper motion of $\sgr$ \citepalias{reid_2020}, these values could be used to reduce the problem to a one-dimensional fit to $\omegab$. To test the effect of including this constraint on $\vphisun$ we repeat the fiducial analysis including the prior $\pi\left(\vphisun\right) \sim \fN\left(250.63,\,0.42\right)$. We then find $\omegab=\psVphiPrior\kmskpc$ and $\vphisun=\vtVphiPrior\kms$, both statistically consistent with the case when no prior is applied.

We alternatively include a prior on the value of $\omegab$ derived from the rotation curve of the models obtained by \citetalias{portail_2017a}. The premise is that, while the models are optimised to fit the bulge data, their rotation curves cannot vary too far from the constraints placed by, for example, \citet{eilers_2019} \& \citet{reid_2019}. We only consider $\vcirc$ data in the range $5<R_{xy}\,\kpcB<6$ as further inwards the assumption of circular motion fails due to the presence of the bar and in the range range $6<R_{xy}\,\kpcB<8$ the models were already fit to the data of \citet{sofue_2009}. Assuming Gaussian error bars the prior is given by,
\begin{equation}
    \log_e\left(
    \pi\left(\omegab|\vcirc\right)
    \right)
    =
    -\frac{1}{2}\sum_i \left[
    \left(\frac{v_{m,\,i} - v_{d,\,i}}{\delta_{d,\,i}}\right)^2 + 2\pi {\delta_{d,\,i}}^2
    \right],
\end{equation}
where $v_{m,\,i}$ ($v_{d,\,i}$) represents the model (data) $\vcirc$ at the $i^\mathrm{th}$ $\ro$ value, and $\delta_{d,\,i}$ represents the corresponding error on the data. 
The measured values of both parameters are given in \cref{c22a:tab:results} and show minor (negligible compared to systematic error) deviations compared to the fiducial case.

We conclude that the gVIRAC data are sufficiently constraining in their own right to provide complementary constraints of the two parameters, independent of previous measurements and deviations of the models from $\vcirc$ measurements just beyond the bar region.


\section{Testing For Systematic Effects}\label{c22a:sec:testing}

In \cref{c22a:sec:errors} we present a comprehensive analysis of the error sources in our measurement. In this section we consider global systematic effects that cannot be accounted for on a voxel by voxel basis.

\subsection{Vary \texorpdfstring{$f_\mathrm{RC\&B}$}{fracFCandB} Requirement}\label{c22a:subsec:test_frcb}

We expect that the adopted Red Clump \& Bump fraction ($f_\mathrm{RC\&B}$, see \cref{c22a:subsec:RCBfrac}) should impact the final results we obtain. 
To quantify this we vary the cutoff, keeping all other assumptions the same, and repeat the outlier-tolerant analysis as described in \cref{c22a:subsec:outliertol}. We consider $f_\mathrm{RC\&B}$ = 10\%, 20\%, 40\%, and 50\% as discussed in \cref{c22a:subsec:RCBfrac}, see \cref{c22a:fig:RCBs_fractions}.
We find that considering 20\% or 10\% cutoffs leads to progressively larger $\omegab$ estimates. Considering the 40\% case leads to a slight decrease, $=-0.1\kmskpc$ from fiducial, while for the 50\% case the value increases up to $\omegab=\psRCandBfifty\kmskpc$; an increase of $\approx1.1\kmskpc$ from fiducial.
For the azimuthal solar velocity we see a minimum value $\approx0.3\kms$ smaller than fiducial for the 40\% case but this rises to $\approx 0.9\kms$ larger for the 50\% case. This sudden rise could be caused by either the effective removal of some systematic effect or the relative lack of data reducing the accuracy of the measurement. As the cutoff fraction increases, the error on the fitted parameters also increases.

\begin{figure}
	\ifTHESIS
	    \includegraphics[width=\columnwidth]{Figures_c22a/vary_FrcAb_LFalpha_ForThesis.pdf}
	\else
	    \includegraphics[width=\columnwidth]{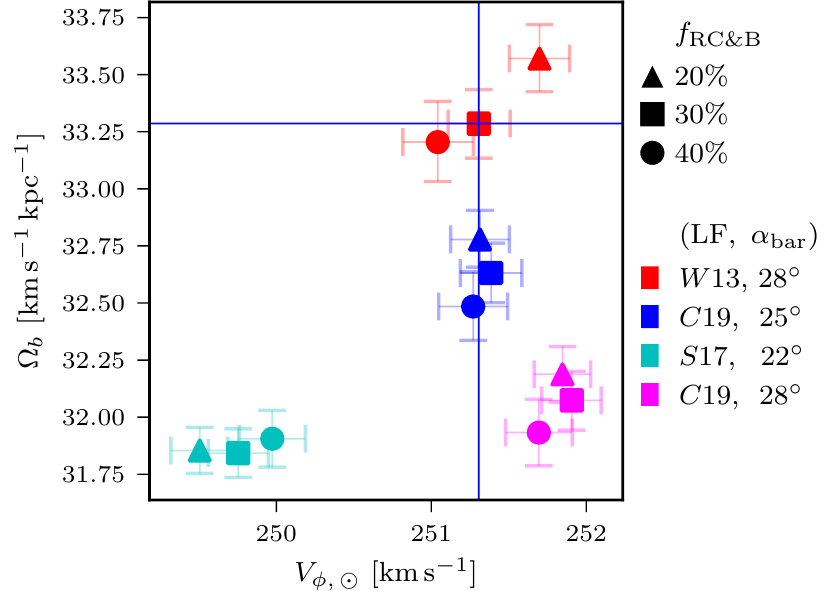}
	\fi
    \caption[Systematic effect of changing (LF, $\bangle$) combination on the derived $\omegab$ and $\vphisun$.]{
    Plot showing the effect of changing the synth-LF and bar angle, $\bangle$, used when predicting the gVIRAC data from the \citetalias{portail_2017a} M2M models. Different $\frcb$ criteria are compared as denoted by different marker shapes and the different (synth-LF, $\bangle$) assumptions are plotted in different colours. The blue lines indicate the result for our fiducial assumptions.
    }
    \label{c22a:fig:vary_LF_alpha}
\end{figure}

We include a contribution to the overall uncertainty equal to the maximum absolute deviation, averaging deviations over (synth-LF, $\bangle$) combinations, from the fiducial value for either the 20\% mask or the 40\% mask. A comparison between the 40\%, 30\%, and 20\% results, for different (synth-LF, $\bangle$) combinations, is shown in \cref{c22a:fig:vary_LF_alpha}. We do not use the more extreme possibilities as the error should represent a reasonable change as opposed to an extreme one. This results in an error component of $\pm\psdeltaF\kmskpc$ for $\omegab$ and $\pm\vtdeltaF\kms$ for $\vphisun$.

\subsection{Vary synth-LF and \texorpdfstring{$\bangle$}{alpha}}\label{c22a:subsec:test_LF_alpha}

Our fiducial assumption is that the (\citetalias{wegg_2013} synth-LF, $\alpha=28\dg$) is a suitable representation of the absolute magnitude distribution in the bulge/bar region; the models are fit to 3D RC density measurements obtained by deconvolving the VVV LOS obs-LFs with the \citetalias{wegg_2013} synth-LF, see \cref{c22a:subsec:predicting_gVIRAC}.
We do indeed find that this combination provides the optimal match to the gVIRAC data of the three that we consider.
However, different studies have predicted different synth-LFs \citepalias[e.g.][]{simion_2017,clarke_2019}, and measurements of the bar angle are correlated to the choice of synth-LF as described in \cref{c22a:subsubsec:lf_bangle_error}.
We therefore treat the choice of synth-LF and $\bangle$ as a coupled system. 
We consider three cases to compare to the fiducial case, (\citetalias{wegg_2013}, $\bangle=28\dg$). The first two cases are discussed in \cref{c22a:subsubsec:lf_bangle_error}: (\citetalias{clarke_2019}, $25^\circ$) and (\citetalias{simion_2017}, $22^\circ$). The final combination we consider, (\citetalias{clarke_2019}, $28^\circ$), tests how the result changes if we do not account for the coupling effect.

The results, for various $f_\mathrm{RC\&B}$ masks, are shown in \cref{c22a:fig:vary_LF_alpha}.
The largest difference occurs for (\citetalias{simion_2017}, $22^\circ$) for which we see average differences of $\Delta\omegab=\psdeltaLFangle\kmskpc$ and $\Delta\vphisun=\vtdeltaLFangle\kms$ compared to the fiducial case. We take these values as the contribution to the overall error as the most conservative estimate. 
The difference between (\citetalias{clarke_2019}, $25\dg$) and fiducial is smaller that the difference for the non-coupled, (\citetalias{clarke_2019}, $28\dg$), case demonstrating the coupling effect between the two parameters.

\subsection{Spiral Structure}

\begin{figure}
	\ifTHESIS
	    \includegraphics[width=\columnwidth]{Figures_c22a/SpiralStructureVIRAC_streamlined_ForThesis.pdf}
	\else
	    \includegraphics[width=\columnwidth]{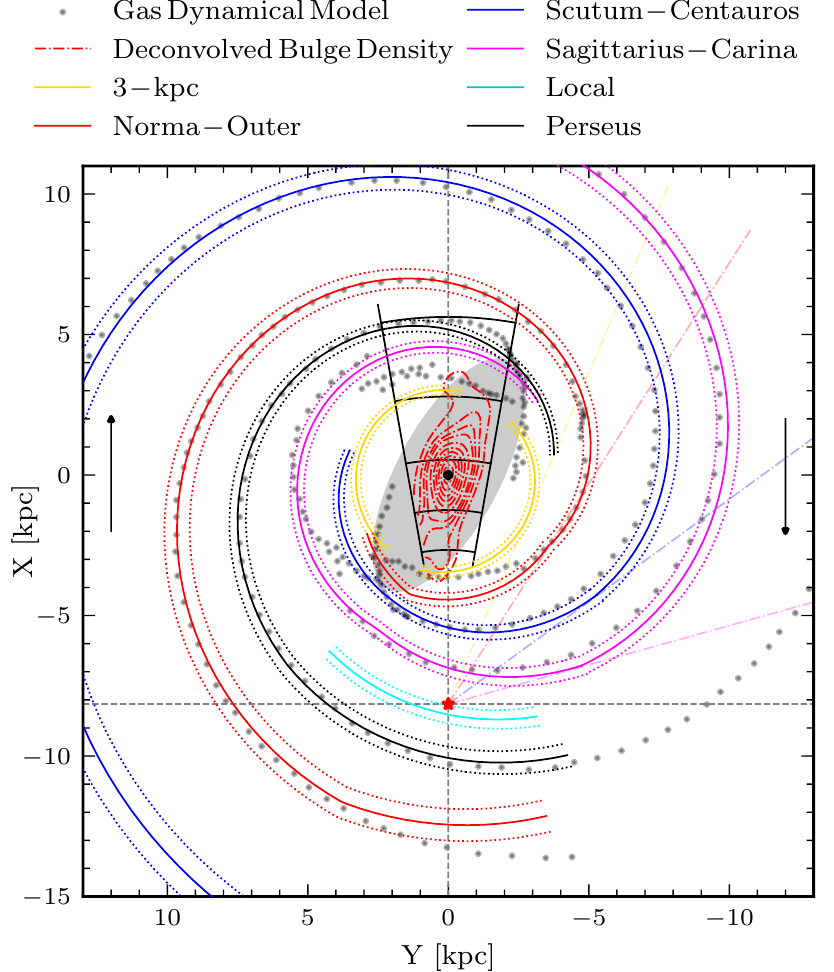}
	\fi
    \caption[Schematic showing the impact of spiral structure on the gVIRAC data.]{
    Face-on map of the MW bar/bulge and spiral arms illustrating results from several studies.
    The red (black) dot shows the location of the sun (Sgr A$^\star$).
    The grey ellipse shows the location and orientation of the Galactic bar as described by \citet{wegg_2015} (half-length=$4.6\kpc$, axis ratio $q=0.4$). 
    The black grid shows the view of the gVIRAC survey in the bulge region; the horizontal lines mark the distance at which a $M_{K_{s0},\,\mathrm{RC}}=-1.694\magn$ RC star would be observed for apparent magnitudes 12.0, 12.5, 13.0, 13.5, and 14.0 $\magn$ {\citepalias{clarke_2019}}.
    \textit{Data}: 
    \begin{inparaenum}
    \protect\item Gas dynamical model \citet{li_zhi_2016},
    \protect\item Deconvolved bulge density \citet{paterson_2020}, and
    \protect\item Spiral arm fits \citet{reid_2019},
    \end{inparaenum}
    as indicated on the figure.
    }
    \label{c22a:fig:spirals}
\end{figure}

There is mounting evidence that the inner MW spiral arms extend inside corotation, perhaps connecting to the ends of the bar, and may even extend within the bar radius \citep[e.g.][]{reid_2019,shen_2020}.
\cref{c22a:fig:spirals} shows a collection of results from various studies aiming to constrain global spiral structure. 
The shaded ellipse shows the location of the long bar \citepalias{wegg_2015}. The black grid shows the gVIRAC viewing area (the horizontal rungs correspond to magnitude intervals for a $M_{K_{s0},\,\mathrm{RC}}=-1.694\magn$ star, see caption).
The grey dots show the location of spiral arms in the gas dynamics simulations of \citet{li_zhi_2016}, the dot-dash red curves show the contours of deconvolved bulge density determined by \citet{paterson_2020}, and the curved arcs are the spiral arm fits computed by \citet{reid_2019} (the faint coloured lines guide the eye to the tangent points of the spirals).

The gas dynamics studies of \citet{li_zhi_2016,li_zhi_2022} found an elliptical structure in the gas which possibly corresponds to the quasi-circular 3-kpc arm found by \citet{reid_2019}. 
As can be seen in \cref{c22a:fig:spirals} the 3-kpc ring can feasibly contaminate the gVIRAC data on the near side and the far side could be contaminated by the 3-kpc, Sagittarius-Carina, and Perseus arm at all longitudes. In addition we see the \citet{paterson_2020} contours show a twisting at the ends which could be related to the 3-kpc arms.

Thus \cref{c22a:fig:spirals} suggests the possibility that the spiral arms overlap with some of the region observed by gVIRAC.
Most foreground stars, i.e. in the Sagittarius-Carina or Scutum-Centauros arms, should have been removed by our colour selection, see \cref{c22a:subsec:gVIRAC}, however it is possible some contamination resides within the gVIRAC RC\&B sample from the 3-kpc arm. At fainter magnitudes if the spiral arms have developed any RGB stars then these will affect the measured kinematics, especially where the bar is relatively less dominant. 
As the models are not capable of capturing the effect of (likely time-evolving) spiral arms, we implement two checks, in the form of additional voxelwise masks, to access the impact spiral structure could have on the final result.

The first is defined by the grey shaded ellipse from \citet{wegg_2015}; any voxel falling outside this boundary is discarded. This amounts to a cut in magnitude, and thus distance given the standard candle nature of RC stars, and should remove all regions in which spiral arms contribute and the kinematics are not necessarily bar dominated. The second, stricter, mask is essentially the same in approach but we use the outermost \citet{paterson_2020} contour which does not show any bending at the end. We refer to these masks as Mask-W15, and Mask-P20 respectively.
Applying these voxelwise masks to the gVIRAC data, and then applying the outlier-tolerant method, we find $\omegab=\psWellipseSpiral\,(\psPcontourSpiral)\kmskpc$ and $\vphisun=\vtWellipseSpiral\,(\vtPcontourSpiral)\kms$ for Mask-W15 (Mask-P20) (results quoted in \cref{c22a:tab:results}).
Mask-P20, implemented to entirely eliminate the effects of spiral structure, results in the maximum difference, relative to the fiducial value, of $\psdeltaSpiral\kmskpc$ for $\omegab$, and $\vtdeltaSpiral\kms$ for $\vphisun$.
This deviation, while small (see \cref{c22a:tab:errors}), is significant compared to the fiducial statistical error, demonstrating that perturbing effects from spiral arms could significantly affect the inferred pattern speed. We thus include a contribution to the overall error, see \cref{c22a:tab:errors}, however the measured $\omegab$ remains a robust bulge/inner bar property given the size of the effect, $<1\kmskpc$.

\subsection{Final Measured Values \& Composite Errors}

In \cref{c22a:tab:errors} we provide a summary of the contributions to the total error from each source of systematic uncertainty. Adding all the different error contributions in quadrature we arrive at our final values: $\omegab=\psfinal\kmskpc$, and $\vphisun=\vtfinal\kms$ where the error in both parameters is dominated by the (synth-LF, $\bangle$) choice.

\begin{table}
\centering
 \caption[Summary of contributions to the overall error together with final results.]{Breakdown of the contribution to the overall error on pattern speed and azimuthal solar velocity from the various tests we have performed. We then give the final values we are reporting with errors (rounded-up) determined in quadrature.}
 \label{c22a:tab:errors}
 \begin{tabular}{l@{\hskip 20pt}cc}
  \hline\\[-5pt]
  Method  &$\omegab$ $\mathrm{[\!\!\kmskpc]}$ & $\vphisun$ $\mathrm{[\!\!\kms]}$ \\[2pt]
  \hline\\[-5pt]
   Fiducial Error               & $\pm\psfiducialerror$ & $\pm\vtfiducialerror$ \\[2pt]
   Effect of $\dpml$ data       & $\pm\psdeltaDisp$     & $\pm\vtdeltaDisp$     \\[2pt]
   Vary $f_\mathrm{RC\&B}$ Mask & $\pm\psdeltaF$        & $\pm\vtdeltaF$        \\[2pt]
   Vary LF \& $\bangle$         & $\pm\psdeltaLFangle$  & $\pm\vtdeltaLFangle$  \\[2pt]
   Spiral Structure             & $\pm\psdeltaSpiral$   & $\pm\vtdeltaSpiral$  \\[2pt]
   \hline\\[-5pt]
   \hfill  & $\psfinal$ & $\vtfinal$\\[2pt]
  \hline
 \end{tabular}
\end{table}

\subsection{Partial Data; Many-Minima Approach}\label{c22a:subsec:many_min}

\begin{figure}
	\ifTHESIS
	    \includegraphics[width=\columnwidth]{Figures_c22a/manyMinima_FrcAb_30_W13_ForThesis.pdf}
	\else 
	    \includegraphics[width=\columnwidth]{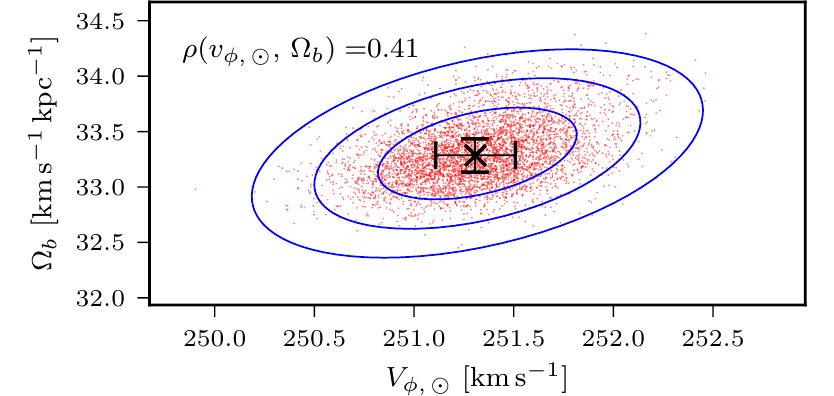}
	\fi
    \caption[Assessment of correlations in our measurement by considering random sub-samples of the data.]{
    Results of the many-minima analysis; we locate the maximum-likelihood point for 5000 25\% random samplings (red dots) of the $\frcb=30\%$ kinematic data comprising 1708 $\mpml$ measurements. The blue ellipses show the 1,2, and 3 $\sigma$ contours and the black errorbar shows the fiducial result using the full sample.
    }
    \label{c22a:fig:many_minima}
\end{figure}

The outlier-tolerant approach, as described in \cref{c22a:subsec:outliertol}, determines the best fitting region of parameter space from the data, models, and errors.  
Some of the voxels are affected by unknown systematic effects, which result in larger model-to-data errors than accounted for in the error analysis, see \cref{c22a:fig:voxelwise_logL_map}.
This could shift the best-fit parameter region away from the true values as the larger errors have disproportionate weights in the likelihood evaluation.
The outlier-tolerant approach, see \cref{c22a:subsec:outliertol}, is only able to approximately account for such systematics.

We thus use a \textit{many-minima} method as an additional test for unknown systematic effects on our results. The premise is simple; we randomly sample voxels, without replacement, from the kinematic data until we have 25\% of the overall sample. 
We take $25\%$ so that a given realisation could be realistically expected to only contain points for which the error is well defined by the analysis in \cref{c22a:sec:errors} while not being so low that the uncertainty on the fitted parameters is overly increased due to loss of constraining power. 
For reference the overall sample in the $\frcb=30\%$ case contains 1708 $\mpml$ measurements.
We then construct the posterior surface and locate the best fitting point. Repeating this process many times provides a 2-dimensional distribution of best-fit points whose distribution in parameter space allows us to access the effect of spurious voxels.

The results of the many-minima analysis are shown in \cref{c22a:fig:many_minima}. The black errorbar shows the location of the fiducial result. The red dots show the best-fit locations for 5000 realisations of the 25\% random sampling and the blue ellipses show the 1, 2, and 3 $\sigma$ regions determined by ellipse fitting to the distribution. 
Because the distribution of the minima scatters evenly around the best-fit value for all data, we conclude that the best-fit result is not significantly biased by the poorly fit voxels.
As expected, the  many-minima $1\sigma$ uncertainty region is larger than that of the fiducial outlier-tolerant result, given that only a quarter of the data is used. There is a correlation between $\vphisun$ and $\omegab$ seen in the many-minima trials but the moderate correlation coefficient $\rho_{\vphisun}^{\omegab}=\manyminCorr$ suggests that the constraints on each parameter are approximately independent.

\subsection{Considering only \texorpdfstring{$\pm l$}{+l} data}

Using a modified form of the \citet{tremaine_1984a} (TW) method to analyse the VIRACv1 proper motions, \citet{sanders_2019b} determined $\omegab=41\pm3\kmskpc$.
This measurement however was restricted to $+l$ data only, as they required it to be consistent with the solar reflex velocity obtained from the proper motion of $\sgr$ \citep{reid_2004} with $\ro=8.12\kpc$.
Relaxing the longitude constraint they obtain $\omegab=31\pm1\kmskpc$ suggesting that the TW method is highly sensitive to systematic effects.

Motivated by this disparity we also evaluate the maximum-likelihood region using only the $(+l,\,\pm b)$ data. For this to be bounded within the model grid, we need, in this case, to additionally exclude the two most in-plane latitude slices in \cref{c22a:fig:voxelwise_logL_map}, avoiding the regions of systematically more negative $\logl$. Using only the $+l$ data results in a small shift in both fitted parameters $(\Delta\vphisun\approx+1.5\kms,\,\Delta\omegab\approx+1.1\kmskpc)$; see \cref{c22a:tab:results}. We conclude that our approach is clearly not subject to such large systematic errors as the TW method.

A similar analysis on the $(-l,\,\pm b)$ side, considering all available data, finds similarly small deviations from the overall result, $(\Delta\vphisun\approx-1.3\kms,\,\Delta\omegab\approx+1.2\kmskpc)$; see \cref{c22a:tab:results}. 
Comparing these results, one may wonder why we find $\omegab\approx34.5\kmskpc$ for each side separately while when using both sides we obtain $\omegab\approx33.3\kmskpc$.
Consider two patches of stars at distances $\Delta X=\pm3\kpc$ from the centre along the bar's major axis and how their kinematics change for small variations, $\Delta\omegab$ and $\Delta\vphisun$. 
For a nearly end-on bar, and to first order, the $v_l$-velocities change by $\Delta v_l\simeq\Delta\omegab\Delta X-\Delta\vphisun$. 
On the near side of the bar ($l>0\dg$ \& $\Delta X=+3\kpc$), if we consider $\Delta\omegab=+0.5\kmskpc$ and $\Delta\vphisun=+1.5\kms$, comparable to those seen between the overall result and the $\pm l$ results, we see $\Delta v_l\simeq (+0.5)(+3)-(+1.5)\simeq 0$; increasing (decreasing) $\omegab$ cancels the variation in $v_l$ due to a suitable increase (decrease) in $\vphisun$.
Conversely for $l<0\dg$ \& $\Delta X=-3\kpc$, if we consider $\Delta\omegab=+0.5\kmskpc$ and $\Delta\vphisun=-1.5\kms$,
we see $\Delta v_l\simeq (+0.5)(-3)-(-1.5)\simeq 0$; increasing (decreasing) $\omegab$ cancels the effect of a suitable decrease (increase) in $\vphisun$.
This simple argument reproduces the sense of how the $\pm l$ results deviate from the full model, and indicates that pattern speed determinations based on only one side of the bar are more vulnerable to such degeneracies than models of the data over the full longitude range.


\section{Resonant Radii in the Disk}\label{c22a:sec:corrotation}

\begin{figure*}
    \ifTHESIS
        \includegraphics[width=\textwidth]{Figures_c22a/corotation_and_OLR_result_plot_E2019_ForThesis.pdf}
    \else
        \includegraphics[width=\textwidth]{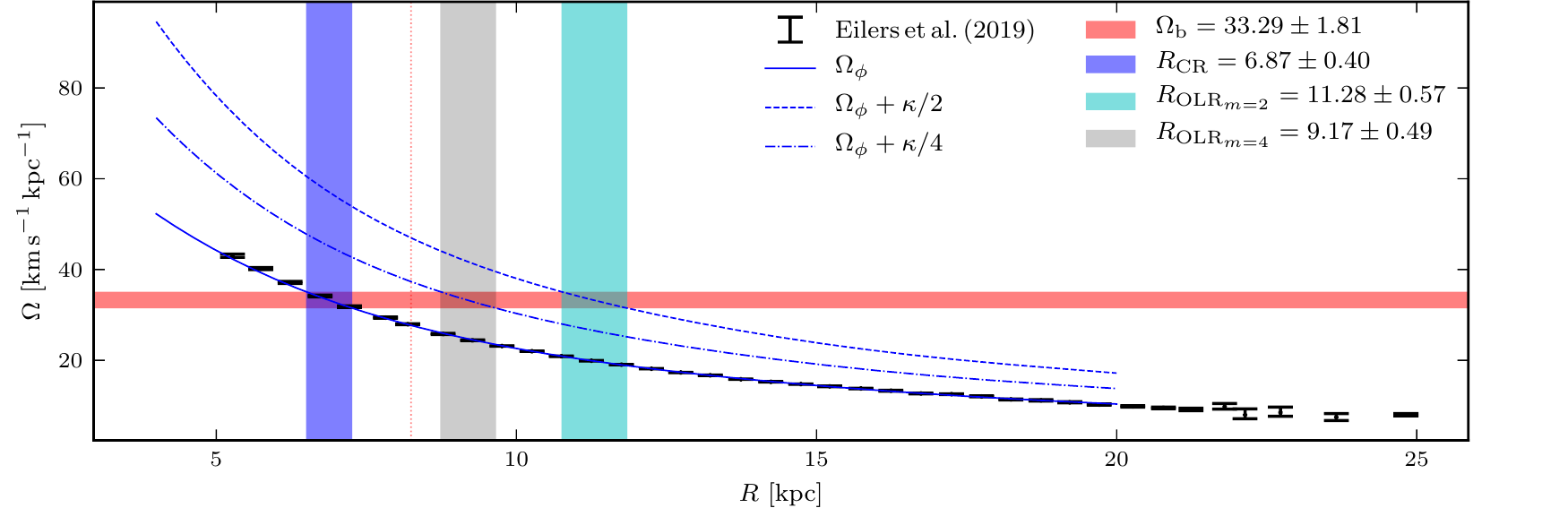}
    \fi
    \caption[Computation of resonant radii by combining our $\omegab$ measurement with the \citet{eilers_2019} rotation curve.]{
    Illustration of the approach taken to estimate the corotation and OLR radii. The data points correspond to the \citet{eilers_2019} rotation curve however we also consider the rotation curve data from \citet{reid_2019}. The solid blue line shows the spline fit to the $\Omega_\phi(R) = \vcirc / R$ while the dashed (dash-dot) blue lines show the $\Omega_\phi(R)$ curve plus the $\kappa/2$ ($\kappa/4$) curves which are used to determine the m=2 (m=4) OLR distance.
    The $\omegab$ measurement made in this paper is outlined by the horizontal red shaded region. The blue vertical shaded region indicates the $\Rcr$ measurement, the cyan shaded region indicates the $\Rolr$ measurement, and the shaded grey region shows the location of the higher order ${\Rolr}_{m=4}$ measurement.
    The vertical red dotted line denotes the \citetalias{gravity_2020} measurement of $\ro$.
    }
    \label{c22a:fig:corrotation_olr}
\end{figure*}

The bar corotation radius, $R_{\mathrm{CR}}$, and outer Lindblad resonance (OLR) radius, $R_{\mathrm{OLR}}$, are key quantities in understanding the MW.
They drive resonances in the disk that produce stellar density features in the SNd as discussed in the introduction.

Resonances occur where there are integer values of $l$ and $m$ that provide solutions to 
\begin{linenomath}\begin{equation}
    m \left( \omegab - \omega_\phi \right) = l \omega_{R},
\end{equation}\end{linenomath}
where $\omegab$ is the bar pattern speed, $\omega_\phi$ is the azimuthal orbital frequency, and $\omega_{R}$ is the radial orbital frequency \citep[p. 188-191]{galactic_dynamics}.
For a nearly circular orbit we can equate $\omega_\phi$ to the circular orbital frequency, $\Omega_\phi\left( R \right)$, and $\omega_{R}$ to the epicyclic frequency, $\kappa\left( R \right)$.
Corotation occurs at $l=0$ and $m=1$ where the star orbits with the bar. The Lindblad resonances occur where $l= \pm 1$ and $m=2$ with $l=+1$ defining the OLR.

We now use our measurement of $\omegab$ to compute estimates of $R_\mathrm{CR}$ and $R_{\mathrm{OLR}}$.
We consider two rotation curves \citep{eilers_2019,reid_2019} which correspond to slightly different circular velocities, ($229.0\pm0.2$, $236\pm7$)$\kms$, and peculiar velocities at the position of the sun.
We use these curves, rather than the models' own rotation curves, as the model rotation curves are only constrained by the dynamics in the bulge region and the \citet{sofue_2009} data for $R_\textrm{GC}=6$-8 kpc, while at intermediate radii and beyond $R_0$ they include a parametric model for the dark matter halo. Therefore while it is possible to measure corotation from the models (as was done in \citetalias{portail_2017a}), they do not reliably constrain the OLR.

We fit a smoothed spline to the $\Omega_\phi = \vcirc/R$ data such that the derivative is also smooth. 
All resonant radii, and corresponding errors, are determined using an iterative numerical bi-section approach.
The corotation radius is determined by locating the distance at which $\Omega_\phi(R) = \omegab$, and the OLR radius is obtained by solving $\omegab = \Omega_\phi\left( R \right) + \kappa\left(R\right)/2$ (see \cref{c22a:fig:corrotation_olr}).
The measured values, for both rotation curves, are given in \cref{c22a:tab:resonant_radii}.
Corotation is found at $\approx6.5<R_\mathrm{CR}\,\,\kpcB<7.5$, and the OLR at $\approx10.7<R_\mathrm{OLR}\,\,\kpcB<12.4$, depending on the assumed rotation curve.
We also find the $m=4$, higher-order OLR distance to be at $8.7 < R_{\mathrm{OLR},\,m=4} \,\, \kpcB < 10.0$, close to the solar radius.

\begin{table}
\centering
 \caption[Tabulation of resonant radii measurements.]{Radii of corotation, OLR (m=2), and the higher order, m=4, OLR for the \citet{eilers_2019} and \citet{reid_2019} rotation curves. All units are in kpc.}
 \label{c22a:tab:resonant_radii}
 \begin{tabular}{l@{\hskip 20pt}cc}
  \hline\\[-5pt]
    & \citet{eilers_2019} & \citet{reid_2019} \\[2pt]
  \hline\\[-5pt]
   Corotation &  $\CoRotEilers$ &  $\CoRotReid$  \\[2pt]
   OLR m=2    & $\OLRaEilers$ & $\OLRaReid$  \\[2pt]
   OLR m=4    &  $\OLRbEilers$ &  $\OLRbReid$  \\[2pt]
  \hline
 \end{tabular}
\end{table}


\section{Discussion}\label{c22a:sec:discussion}

\begin{figure}
    \ifTHESIS
    	\includegraphics[width=\columnwidth]{Figures_c22a/schematic_measurement_ForThesis.pdf}
    \else
    	\includegraphics[width=\columnwidth]{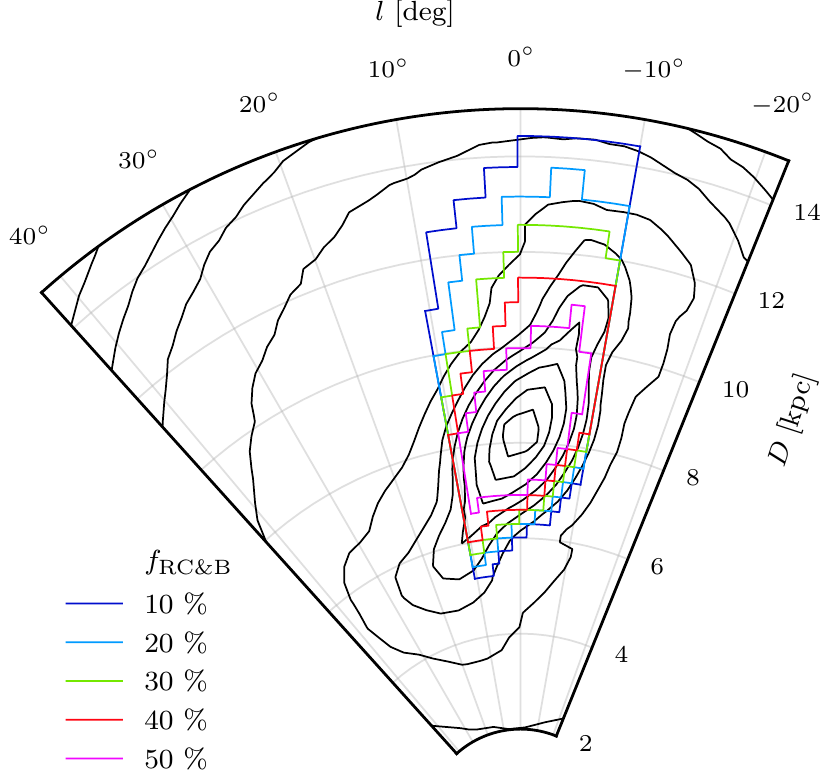}
    \fi
    \caption[Schematic showing the region in which we measure $\omegab$ \& $\vphisun$.]{
    Schematic showing the region in which we measure $\omegab$; different $\frcb$ masks are outlined by the coloured regions and superimposed on top of the bulge density contours computed from \citetalias[the $\omegab=37.5\kmskpc$ model of][]{portail_2017a}. Distances are computed by converting magnitudes assuming our fiducial RC magnitude, $M_{\ks,\,\mathrm{RC}}=-1.694\magn$. The masks demonstrate that we are measuring the pattern speed of the b/p bulge with some contribution from the outer bulge/long-bar region for $\frcb \lesssim 30\%$.
    }
    \label{c22a:fig:measurement_location}
\end{figure}

\begin{figure}
    \ifTHESIS
	    \includegraphics[width=\columnwidth]{Figures_c22a/omegab__UND__vphisun_history_ForThesis.pdf}
    \else
	    \includegraphics[width=\columnwidth]{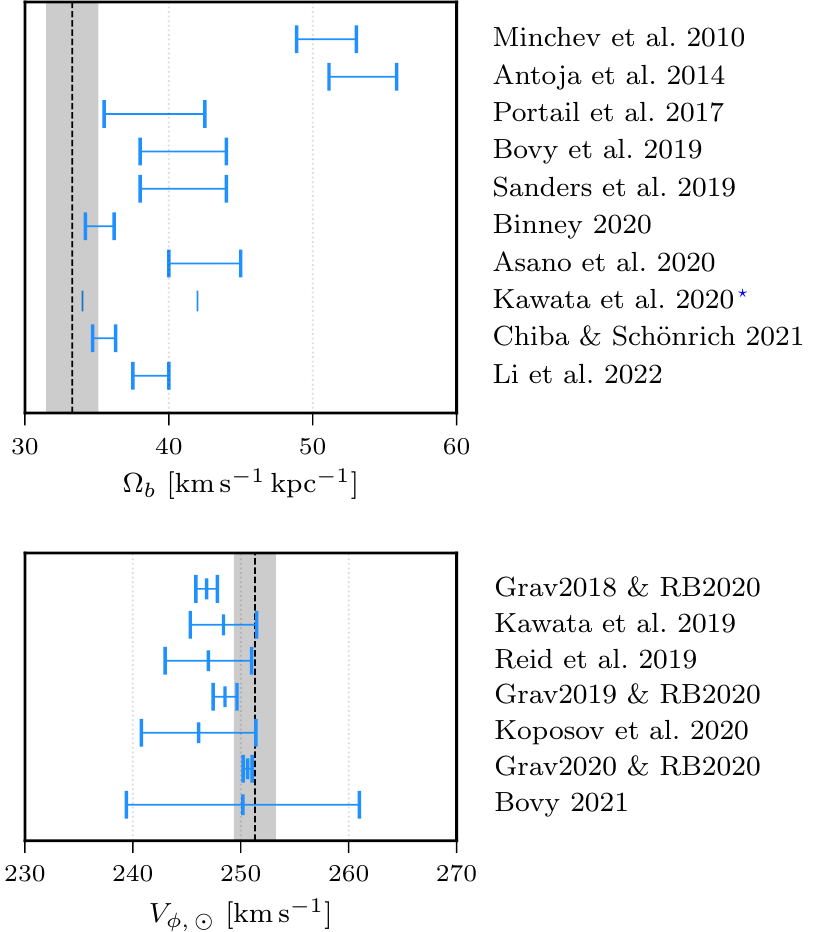}
	\fi
    \caption[Previous literature measurements of $\omegab$ \$ $\vphisun$.]{
    The results of this work are shown as the vertical line and the shaded region gives the error bar. 
    \textit{Top:} Compilation of pattern speed measurements from the literature. 
    \textcolor{blue}{$^\star$} \citet{kawata_2021} found that two values of $\omegab$ could reproduce the local solar velocity substructures equally well.
    \textit{Bottom:}
    Compilation of previous $\vphisun$ measurements from the literature. 
    }
    \label{c22a:fig:omega_vphisun_history}
\end{figure}

We have measured the Milky Way bar's pattern speed to be $\omegab=\psfinal\kmskpc$ by comparing VIRAC $\mpml$ and $\dpml$ proper motion data to a grid of M2M models from \citetalias{portail_2017a}. 
\cref{c22a:fig:measurement_location} shows a schematic of the measurement area superimposed on the bulge density contours from the $\omegab=37.5\kmskpc$ M2M model. The outlined regions show the coverage of the five $\frcb$ masks considered in this work; the magnitude limits have been converted to distance following $m_\ks - M_\ks = 5\log_\mathrm{10}{\left(D/10\pc\right)}$ and assuming $M_\ks = M_{\ks,\,\mathrm{RC}} = -1.694 \magn$ (\cref{c22a:subsec:predicting_gVIRAC}).
The regions demonstrate that $\frcb=50\%$ effectively samples the b/p bulge region while conversely the $\frcb=10\%$ mask extends along the long-bar and includes regions of the outer bulge and inner disk. As such we primarily measure the pattern speed of the inner bar and bulge region. The remarkable agreement between the different $\frcb$ results, see \cref{c22a:tab:results}, indicates our results are consistent with uniform solid body rotation; we find no evidence for a systematic variation with scale, or that the b/p bulge and the long-bar rotate with different pattern speeds.

In \cref{c22a:fig:omega_vphisun_history} we show previous literature estimates of $\omegab$ (top) and $\vphisun$ (bottom).
For comparison the estimates made in this paper are shown by the vertical black line and the error bar by the shaded grey region.
Our $\omegab$ measurement is slightly smaller than a number of recent measurements; $\omegab=36.0\pm1.0 \gyr^{-1} = 35.2\pm1.0 \kmskpc$ \citep[orbit trapping by bar resonances,][]{binney_2020} and $\omegab=35.5\pm0.8\kmskpc$ \citep[mean metallicity gradient of stars trapped by the resonance of a decelerating bar,][]{chiba_2021b}, despite being based on completely independent data (bulge vs local disk kinematics).
We are also in excellent agreement with one of the two values favoured by \citet{kawata_2021}, $\omegab=34\kmskpc$, who considered multiple higher-order bar resonances to match local velocity substructure.
These complementary analyses thus result in a highly consistent measurement for $\omegab$ considering data from the bulge/bar region out to the bar resonances in the SNd.

Furthermore our $\vphisun$ measurement is within $\approx1\sigma$, at the high end, of a large body of previous work that generally agrees on $\vphisun\approx250\kms$. Note the excellent consistency with the value of $\vphisun$ derived when combining the \citetalias{gravity_2020} and \citetalias{reid_2020} measurements; there is no suggestion that $\sgr$ is not at rest at the centre of the larger bulge structure.

\citet{hilmi_2020} recently demonstrated that galactic bar parameters, such as $\omegab$ and bar length, can fluctuate due to interactions with spiral arms \citep[see also, e.g., ][]{quillen_2011,martinez_valpuesta_2011}. 
In their models they found that the bar length could fluctuate by up to 100\% and $\omegab$ vary by up to $\approx$20\% on a time scale of 60 to 200 Myr.
They then argue that, were $\omegab$ for the MW bar region fluctuating by as much as $20\%$, the recent \citet{bovy_2019,sanders_2019b} \textit{`instantaneous'} measurements would still be consistent with their advocated, \textit{`time-averaged'} $\omegab\sim 50\kmskpc$ \citep[e.g.][]{minchev_2007,antoja_2014}, see \cref{c22a:fig:omega_vphisun_history}.
However our measurement, and those of \citet{binney_2020,chiba_2021b}, would remain inconsistent with this larger value.

The periodic connection and disconnection of the bar and spiral arms observed by \citet{hilmi_2020} also perturbs the corotation resonance. First, the pattern speed $\Omega_b$ of the bar itself varies, accelerating (decelerating) before connecting (disconnecting) to a spiral arm. Second, because the bar and spiral-arm potentials superpose, the potential's average pattern speed $\Omega_{\rm m\star}$ in the resonance region varies when significant spiral arm mass enters into or rearranges near the bar's corotation radius, on dynamical time-scales. In a fixed reference frame rotating with, e.g., the average bar pattern speed this corresponds to time-dependent forces.
These effects would shift the corotation resonance and continuously move stars in and out of the resonance. Because the libration periods of the Lagrange orbits are of order Gyr, phase-dependent perturbations should be visible for a long time. However, in the MW a high degree of phase mixing for these orbits is indicated by the analysis of \citet[][Figs.~4 \& 5 therein]{binney_2020}, arguing against strong bar fluctuations in the MW.

The hypothesis that measurements in the SNd constitute a time-averaged measurement of $\Omega_\mathrm{m\star}$, or $\omegab$, is itself questionable. Assuming $\omegab=35\kmskpc$, the time for one full bar rotation is $\tau_\mathrm{bar}\approx 175\myr$, whereas for $\ro=8.2\kpc$ and $\vcirc(\ro)=230\kms$, the period of a circular orbit at the sun's distance is $\tau_\mathrm{\circ}(\ro)\approx220\myr$. This is only a $\approx 25\%$ difference and suggests that SNd kinematics would also be sensitive to fluctuations in $\omegab$.

A further consideration is the timescale over which bar fluctuations and deceleration occur. \citet{li_zhi_2022}, using modified versions of the M2M bar potentials from \citetalias{portail_2017a}, and including spiral arms, studied hydrodynamical simulations of the gas dynamics in the inner Galaxy. They found their gas reaches quasi steady state on a timescale of $\sim 300 \myr$, longer than the bar fluctuation timescale of $60-200\myr$ found by \citet{hilmi_2020}. Matching their gas flow models to various features in the Galactic $(l,\,v_\mathrm{los})$ diagram, \citet{li_zhi_2022} determine a best pattern speed, $37.5<\omegab\,\,(\!\kmskpc)<40.0$. They argue that their measurement is essentially time averaged because the gas cannot immediately respond to changes to the underlying potential. 
The situation is further complicated when one considers the effects of a decelerating bar. The bar's pattern speeds generally slows down over time due to transfer of angular momentum to the dark matter halo \citep[e.g.][]{weinberg_1985,debattista_2000,valenzuela_2003,martinez_valpuesta_2006,sellwood_2008}.
\citet{chiba_2021a} show that a decelerating bar can explain the structure of the Hercules stream in local velocity and angular momentum space, and is also able to generate similar structures and patterns as seen in local SNd data which are often attributed to resonances of a constant $\omegab$ bar or transient spiral structure.
The inferred bar deceleration rate, $\ChibaResult$ \citep{chiba_2021a}, leads to a change in $\omegab$ by $1.35\kmskpc$ in $300\myr$. When compared to the final result of \citet{li_zhi_2022}, the bar slowdown, combined with the gas' inability to immediately adapt to the slowing potential, could extend their plausible range of $\omegab$ down to $\approx36\kmskpc$, in approximate agreement with the present work. However this is not clear since the results of \citet{li_zhi_2022} are unchanged if they rerun their hydrodynamical simulations with a decelerating bar.

\newcommand{\PortailCR}{ R_\mathrm{CR}=6.1\pm0.5\kpc }
\newcommand{\SandersCR}{ R_\mathrm{CR}=5.7\pm0.4\kpc }

Using our measurement of the bar's pattern speed together with the Galactic rotation curves of \citet{eilers_2019} and \citet{reid_2019}, we infer values $R_\mathrm{CR}=6.5-7.5\kpc$ for the co-rotation radius, and $R_{\mathrm{OLR}}=10.7-12.4\kpc$ for the outer Lindblad resonance radius. These are slightly larger than values quoted recently based on somewhat higher values of $\omegab$ estimated, e.g., from M2M dynamical modelling \citep[][$\PortailCR$]{portail_2017a}, or from the application of the continuity equation to VIRAC and \textit{Gaia} proper motion data \citet[$\SandersCR$]{sanders_2019b}.
The $m=4$, higher-order OLR found with our value of $\omegab$ is at $8.7 < R_{\mathrm{OLR},\,m=4} \,\, \kpcB < 10.0$, making it likely that it too contributes to the complex velocity structure found in the SNd \citep[see also][]{hunt_2018b,kawata_2021}.

As for $\omegab$-independent evidence, \citet{khoperskov_2020} found six arc-like density structures in angular momentum space in spatially homogenized \textit{Gaia} star counts. Of these, they associated one at $\approx 6.2 \kpc$ to orbits near the co-rotation resonance and one at $\approx9\kpc$ to orbits around the OLR. These radii are smaller than the values we determine and it appears plausible that the $9\kpc$ feature is actually associated to the $m=4$ higher order OLR resonance rather than the $m=2$ OLR.
\citet{binney_2020} and \citet{chiba_2021b} infer their preferred values for the pattern speed from matching the bar's co-rotation resonance to the Hercules stream \citep{perez_villegas_2017}.
The OLR is then associated to one of the streams at higher $v_\phi$, plausibly the Sirius stream.

\section{Conclusion}\label{c22a:sec:conclusion}

We have compared distance-resolved VIRAC-\textit{Gaia} (gVIRAC) proper motion data in the Galactic b/p bulge and bar to a grid of M2M models with well defined pattern speeds from P17, to investigate the bar's pattern speed and the solar azimuthal motion.
We have undertaken a comprehensive assessment of the statistical and systematic errors present in our measurements, including spatial variations and magnitude dependence of the correction to the \textit{Gaia} absolute reference frame, the extraction of the RC\&B from the RGB luminosity function, the magnitude-dependent broadening of the RC\&B kinematics due to the VIRAC proper motion errors, and uncertainties due to the M2M modelling.
We use a robust outlier-tolerant statistical approach to quantitatively compare the gVIRAC data to the grid of models and test the systematic effects of varying the assumption of LF, bar angle $\bangle$, RC\&B threshold, and the possible overlap from spiral arms. We include contributions to the final error from these sources.

We find that the best P17 model matches the gVIRAC $\mpml$ data to an \textsc{rms} precision of $<9\kms$ for the fiducial case in which red clump giant stars have a statistical weight of more than $30\%$ in a given voxel. This is despite the fact that the P17 models have not been fit to the gVIRAC data but are based on star-count and LOS velocity data and are used solely to predict the gVIRAC kinematics.

Using the marginalized posterior probability curves, and adding errors from systematic effects in quadrature, we obtain $\omegab=\psfinal\kmskpc$ and $\vphisun=\vtfinal\kms$ which are in excellent agreement with the best recent determinations from solar neighbourhood data.
Combining our $\omegab$ measurement with recent rotation curve determinations we find corotation to be at $\approx7.0\pm0.5\kpc$, the OLR to be at $\approx 11.55 \pm 0.85\kpc$ and the $m=4$ OLR to be at $\approx 9.35\pm0.65\kpc$.

Linking our result with recent measurements of the pattern speed from the Hercules stream (corotation resonance) in the SNd, a self-consistent scenario emerges in which the bar is large and slow (albeit dynamically still relatively fast), with $\omegab\simeq\,35\kmskpc$, based on data both in the bar/bulge and in the SNd.

In future work we shall fit a new generation of M2M models to the gVIRAC data with which to quantitatively explore the dynamics and mass distribution, both baryonic and dark, in the inner Galaxy.

\section*{Acknowledgements}
We gratefully acknowledge the anonymous referee for their helpful comments, Leigh C. Smith for continued advice and support in using the VIRACv1 data, and Shola M. Wylie for useful discussions, which have all led to improvements in the paper.
Based on data products from VVV Survey observations made with the VISTA telescope at the ESO Paranal Observatory under programme ID 179.B-2002.
This work has made use of data from the European Space Agency (ESA) mission
{\it Gaia} (\url{https://www.cosmos.esa.int/gaia}), processed by the {\it Gaia}
Data Processing and Analysis Consortium (DPAC,
\url{https://www.cosmos.esa.int/web/gaia/dpac/consortium}). Funding for the DPAC
has been provided by national institutions, in particular the institutions
participating in the {\it Gaia} Multilateral Agreement.

\section*{Data Availability}

The data underlying this article will be shared on reasonable request to the corresponding author.

\section*{ORCID iDs}

Jonathan Clarke \orcidA{} \href{https://orcid.org/0000-0002-2243-178X}{https://orcid.org/0000-0002-2243-178X}\\
Ortwin Gerhard \orcidB{} \href{https://orcid.org/0000-0003-3333-0033}{https://orcid.org/0000-0003-3333-0033}

\bibliographystyle{mnras}
\bibliography{omega_c22a} 



\appendix
\section{Accounting For Bulge Vertical Metallicity Gradients}\label{c22a:appendix:metallicity}
\begin{figure}
    \centering
    \ifTHESIS
        \includegraphics[width=\columnwidth]{Figures_c22a/w13_R0_z_relation_2magitude_relationship_ForThesis.pdf}
    \else
        \includegraphics[width=\columnwidth]{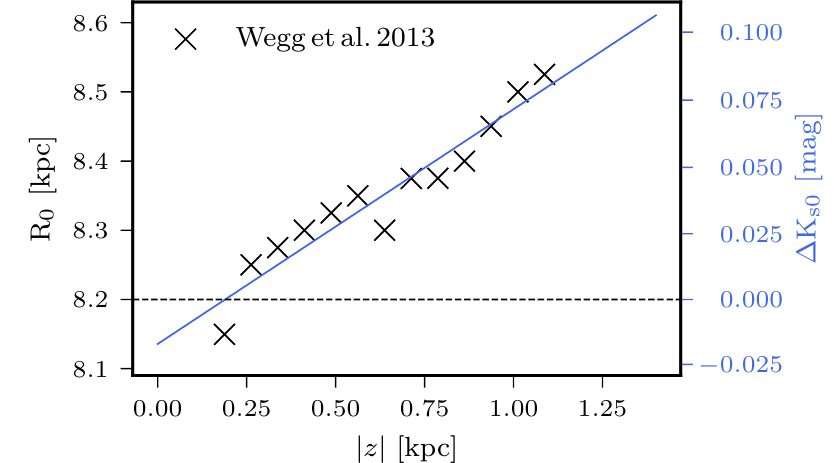}
    \fi
    \caption[
    Relationship between z and derived value for {$\ro$} obtained by {\citetalias[][see their Fig. 10]{wegg_2013}}.
    ]{
    Relationship between z and derived value for $\ro$ obtained by \citetalias[][see their Fig. 10]{wegg_2013}. The black crosses show their data which is consistent with the vertical metallicity gradient shifting $M_{K_{s0},\, \mathrm{RC}}$ to slightly fainter magnitudes with increasing vertical height above the Galactic plane. The blue line shows a linear regression fit to the data points and the blue y axis shows the shift in magnitude equivalent to the difference in distance relative to the fiducial value, $R_0 = 8.2 \kpc$.
    }
    \label{c22a:app:fig:met_grad}
\end{figure}

\cref{c22a:app:fig:met_grad} shows the approach taken to account for the bulge vertical metallicity gradient which, when one assumes a constant $M_{K_{s0},\, \mathrm{RC}}$, manifests as an apparent shift in the distance to the GC. We have taken the data from \citetalias[][(Fig. 10)]{wegg_2013} which shows an apparent difference in the distance to the GC of $\sim 0.4 \kpc$, corresponding to a magnitude difference of $\sim0.1$ mag. This is caused by the vertical metallicity gradient shifting $M_{K_{s0},\, \mathrm{RC}}$ to fainter magnitudes with increasing height; \citet{gonzalez_2013} found a gradient of $0.28\, \mathrm{dex} \, \mathrm{kpc^{-1}}$ which, when combined with $\mathrm{dM_{K_{s0},\,RC}/d([Fe/H]) = 0.275}$ \citep{salaris_2002}, predicts $\Delta M_{K_{s0}} = 0.09 \magn \, \mathrm{kpc}^{-1}$.

To account for the metallicity gradient we fit a straight line to the points using linear regression; we obtain a gradient, $\beta=0.33926$, and intercept, $\alpha=8.13571$. When observing the model we place the sun at $8.2\kpc$ from the centre of the bulge so we take this as the zero point. The effect of the vertical metallicity gradient on the apparent magnitude is then described by,
\begin{equation}
    \Delta \ks = 5 \log_{10}\left( \frac{\beta |z| + \alpha}{8.2} \right),
\end{equation}
which is added to each particles' apparent magnitude as it is observed.

\bsp	
\label{lastpage}
\end{document}